\documentclass[journal]{IEEEtran}
\usepackage{algorithm}
\usepackage{algpseudocode}
\usepackage{amsfonts}
\usepackage{bm}
\usepackage{amsmath}
\usepackage{cite}
\usepackage{float}
\usepackage{amssymb}
\usepackage{enumerate}
\usepackage{color,xcolor}
\usepackage{graphicx}
\usepackage{multirow,tabularx}
\usepackage{subfigure}
\usepackage{hyperref}

\makeatletter

\newcommand{\Rmnum}[1]{\expandafter\@slowromancap\romannumeral #1@}
\makeatother

\ifCLASSINFOpdf

\else

\fi

\PassOptionsToPackage{draft}{hyperref}

\begin{document}

\title{Convergence Analysis and Assurance for Gaussian Message Passing Iterative Detector in Massive MU-MIMO Systems}

\author{\IEEEauthorblockN{Lei Liu, \emph{Student Member, IEEE,} Chau Yuen, \emph{Senior Member, IEEE,}\\ Yong Liang Guan, \emph{Member, IEEE,} Ying Li, \emph{Member, IEEE} and Yuping Su}

\thanks{Lei Liu, Ying Li and  Yuping Su are with the State Key Lab of Integrated Services Networks, Xidian University, Xi'an, 710071, China (e-mail: lliu\_0@stu.xidian.edu.cn, yli@mail.xidian.edu.cn, ypsu@stu.xidian.edu.cn).}
\thanks{Chau Yuen is with the Singapore University of Technology and Design, Singapore (e-mail: yuenchau@sutd.edu.sg).}
\thanks{Yong Liang Guan is with the School of Electrical and Electronic Engineering, Nanyang Technological University, Singapore (e-mail: eylguan@ntu.edu.sg).}
}


\maketitle

\begin{abstract}
This paper considers a low-complexity Gaussian Message Passing Iterative Detection (GMPID) algorithm for massive Multiuser Multiple-Input Multiple-Output (MU-MIMO) system, in which a base station with $M$ antennas serves $K$ Gaussian sources simultaneously. Both $K$ and $M$ are very large numbers, and we consider the cases that $K<M$. The GMPID is a low-complexity message passing algorithm based on a fully connected loopy graph, which is well understood to be not convergent in some cases. As it is hard to analyse the GMPID directly, the large-scale property of the massive MU-MIMO is used to simplify the analysis. Firstly, we prove that the variances of the GMPID definitely converge to the mean square error of Minimum Mean Square Error (MMSE) detection. Secondly, we propose two sufficient conditions that the means of the GMPID converge to those of the MMSE detection. However, the means of GMPID may not converge when $ K/M\geq (\sqrt{2}-1)^2$. Therefore, a new convergent GMPID called SA-GMPID (scale-and-add GMPID) , which converges to the MMSE detection in mean and variance for any $K<M$ and has a faster convergence speed than the GMPID, but has no higher complexity than the GMPID, is proposed. Finally, numerical results are provided to verify the validity and accuracy of the theoretical results.
\end{abstract}


\linespread{1.38}

\IEEEpeerreviewmaketitle
\section{Introduction}
Recent research investigations\cite{Argas2013,Rusek2013,biglieri2007} have shown that Multiuser Multiple-Input and Multiple-Output (MU-MIMO) will play a critical role in the future wireless systems. MU-MIMO has become a key technology for wireless communication standards like IEEE 802.11n, IEEE 802.11ac, WiMAX and Long Term Evolution. More recently, the massive MU-MIMO technology, in which the Base Station (BS) has a very large number of antennas (e.g., hundreds or even more), has attracted more and more attention \cite{Argas2013, Rusek2013,biglieri2007, Hoydis2013, Studer2013, Marzetta2010, Ngo2012, Dai2013, Fernandes2013}. In particular, massive MU-MIMO has been shown to be able to bring significant improvement both in throughput and energy efficiency, and thus meet the growing demands for higher throughput and quality-of-service of the next-generation communication systems\cite{Marzetta2010,Ngo2012,Dai2013}.

The costs of introducing massive MIMO include more physical space at BS, higher complexity, and higher energy consumption for signal processing at the transmitters and receivers\cite{Rusek2013}. Low-complexity uplink signal detection for massive MU-MIMO is hence desirable\cite{Rusek2013}. In the case of linear detection of Gaussian sources in noisy channels, it is well known that Minimum Mean Square Error (MMSE) detection is optimal. However, its computational complexity is high due to the need to perform large matrix inversion \cite{tse2005}. To avoid the matrix inversion, some classical iterative algorithms like Jacobi algorithm, Richardson algorithm, Neumann Series and Gauss-Seidel algorithm may be applied \cite{Bertsekas1989, Henrici1964, Wu2014, Gao20142, Gao2014,Dai20152}. Another promising MU-MIMO detection is a graph-based detection called message passing algorithm (MPA) \cite{ Forney2001, kschischang2001, Loeliger2002, Loeliger2004, Loeliger2006}. It is also linked to the canonical problem of solving systems of linear equations \cite{moallemi2009, Loeliger2002}, which is encountered in many computer science and engineering problems such as signal processing, linear programming, ranking in social networks, support vector machines, etc.\cite{kschischang2001, Loeliger2002,Loeliger2004,Loeliger2006,Bickson2008, bickson2009}. There are two types of MPAs. The first is the Gaussian Belief Propagation (GaBP) algorithm based on a graph that consists of variable nodes \cite{Weiss20011, Weiss20012, malioutov2006, Su2014, moallemi2009}. The second is the Gaussian Message Passing Iterative Detection (GMPID) algorithm based on a pairwise graph that consists of variable nodes and sum nodes \cite{Guo2008, andrea2005, Roy2001, William2009, wu2014, yoon2014, Boutros2002}. Both of them are efficient distributed algorithms for Gaussian graphical models. In particular, GMPID has also been extensively studied for equalization in the inter-symbol interference channel \cite{Guo2008}, and decoding of modern channel codes, such as turbo codes and Low Density Parity Check (LDPC) codes \cite{William2009}.

It has been proved that if the factor graph has a tree structure, the means and variances of the MPA converge to the true marginal means and approximate marginal variances respectively \cite{kschischang2001,Loeliger2004}. However, if the graph has cycles, the MPA may fail to converge. To the best of our knowledge, most previous works of the MPA focus on the convergence of the GaBP algorithm. Three sufficient conditions for the convergence of GaBP in loopy graphs are known: diagonal-dominance \cite{Weiss20011,Weiss20012}, convex decomposition \cite{moallemi2009} and walk-summability \cite{malioutov2006}. Recently, a necessary and sufficient variance convergence condition of GaBP is given in \cite{Su2014}. For the GMPID based on the pairwise graph, a sufficient condition of the mean convergence is given in \cite{Roy2001} and it is shown that 1) the covariance matrices definitely converge, 2) if they converge, the means of GMPID coincide with the true marginal means. However, in this GMPID, posterior density matrices of the sum nodes need to be calculated \cite{Roy2001}, which introduces the matrix inversion operation and leads to a much higher computational complexity during the message updating. In general, the GMPID algorithm has lower computational complexity and better Mean Square Error (MSE) performance than the GaBP algorithm. Actually, the MU-MIMO system can be regarded as a randomly-spread CDMA channel \cite{Boutros2002} by treating the antennas as different time chips. Montanari \cite{andrea2005} has proved that GMPID converges to the optimal MMSE solution for any arbitrarily loaded randomly-spread CDMA system. However, the proof works only for CDMA MIMO system with binary channels. To the best of our knowledge, most previous works focus on the convergence of the MPA based on the graphs that consist of only variable nodes (like GaBP). On the other hand, the convergence of MPA based on pairwise graphs that consist of variable nodes and sum nodes (like GMPID) is far from solved.

In this paper, we analyse the convergence of GMPID and propose a new low-complexity fast-convergence multi-user detector for massive MU-MIMO system with $K$ users and $M$ antennas. Let $\beta  = {K \mathord{\left/
 {\vphantom {K M}} \right.
 \kern-\nulldelimiterspace} M}$ and $\beta<1$. The contributions of this paper are summarized as follows:

 \noindent
\hangafter=1
\setlength{\hangindent}{2em} 1) We prove that the variances of GMPID definitely converge to the MSE of MMSE detection, which gives a simple alternative way to estimate the MSE of the MMSE detector.

 \noindent
\hangafter=1
\setlength{\hangindent}{2em} 2) Two sufficient conditions, which show that the means of GMPID converge to those of the MMSE detector for $\{\beta: 0<\beta<(\sqrt{2}-1)^{2}\}$, are derived.

 \noindent
\hangafter=1
\setlength{\hangindent}{2em} 3) A new fast-convergence detector called SA-GMPID, which converges to the MMSE detection in mean and variance and has a faster convergence speed than the GMPID for any $\{\beta: 0<\beta<1\}$, is proposed.

This paper is organized as follows. In Section II, the MU-MIMO model and MMSE detector are introduced. The GMPID is elaborated in Section III. Section IV presents the proposed fast-convergence detector SA-GMPID. Numerical results are shown in Section V, and we end this paper with conclusions in Section VI.

\section{System Model and MMSE Detector}
In this section, the massive MU-MIMO system model and some preliminaries about the MMSE detection algorithm for the massive MU-MIMO systems are introduced.

\subsection{System Model}

\begin{figure}[ht]
  \centering
  \includegraphics[width=5.5cm]{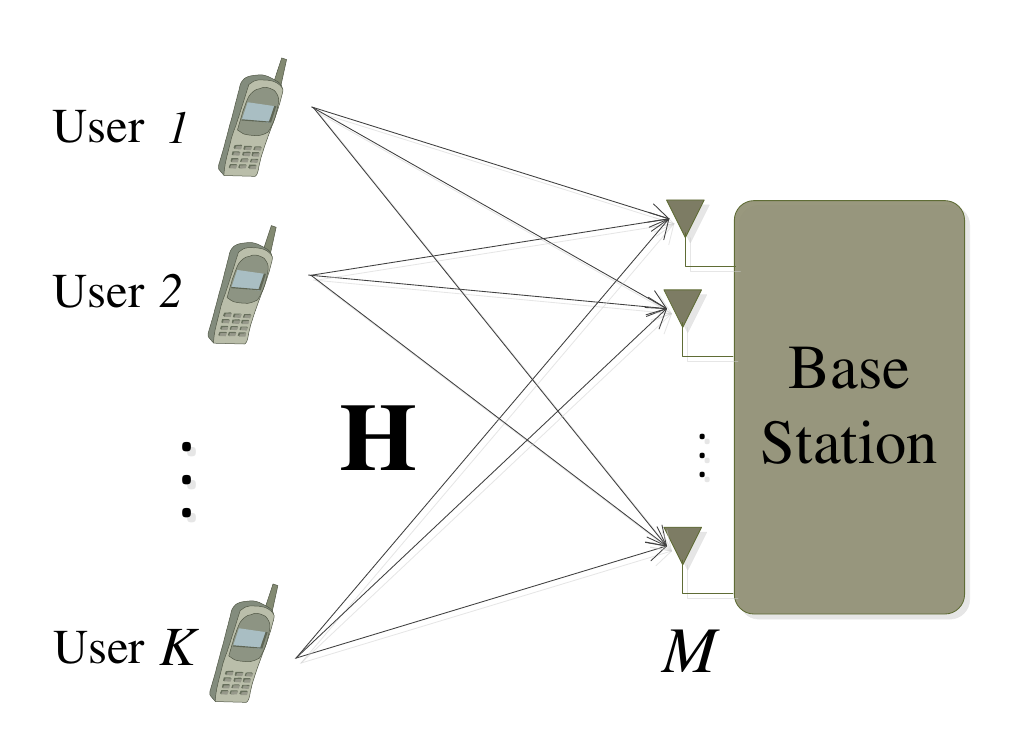}\\
  \caption{$K\times M$ MU-MIMO system model: $K$ autonomous single-antenna terminals communicate with an array of $M$ antennas of the base station.}\label{f1}
\end{figure}
Fig. \ref{f1} shows an uplink MU-MIMO system with $K$ users and one BS with $M$ antennas \cite{Marzetta2010, biglieri2007, Ngo2012, Rusek2013}. For massive MIMO, the $K$ and $M$ are in hundreds or thousands, e.g., $M=600$ and $K=100$. The $M\times1$ received signal vector ${\large{\textbf{\emph{y}}} }$ at the BS can be represented by
\begin{equation}\label{e1}
{\large{\textbf{\emph{y}}} }= \mathbf{H}\textbf{\emph{x}} + \textbf{\emph{n}},
\end{equation}
where $\mathbf{H}$ denotes the $M \times K$ channel matrix, $\textbf{\emph{n}}$ is an $M \times 1$ independent Additive White Gaussian Noise (AWGN) vector, i.e., $\textbf{\emph{n}}\!\!\sim\!\!\mathcal{N}^{\small{M}}(0,\sigma_n^2)$, and $\large\textbf{\emph{x}}$ is the message vector sent by $K$ users. Each component of $\large\textbf{\emph{x}}$ is Gaussian distributed, i.e., ${x_k}\!\!\sim\!\!\mathcal{N}(0,\sigma_{x_k}^2)$ for $k \!\!\in\!\! \{ 1,2,\cdots,K\} $. We assume that the channels only suffer from small-scale fading without large scale fading, in which case $\mathbf{H}$ takes the form of a Rayleigh fading channel matrix whose entries are independently and identically distributed (i.i.d.) Gaussian random variables with zero means and unit variances, i.e., normal distributions $\mathcal{N}^{M\times K}(0,1)$. The task of multi-user detection at the BS is to estimate the transmitted signal vector ${\large{\textbf{\emph{x}}} }$ from the received signal vector ${\large{\textbf{\emph{y}}} }$. In this paper, we assume that the BS knows the $\mathbf{H}$, and we only consider the real MU-MIMO system because the complex case can be easily extended from the real case \cite{Gao2014}.

{\subsection{Gaussian Source Assumption}
In the real communication systems, discrete modulated signals are generally used. However, according to the Shannon theory \cite{Cover2006,Gamal2012}, the capacity of Gaussian channel is achieved by a Gaussian input. Therefore, the independent Gaussian sources assumption is widely used in the design of communication networks \cite{Abramovich2001, Kafedziski2012, Abramovich2007, Liu2012, Matamoros2011, Bao2008}. It means that, in real systems, to achieve a high transmission rate, the distributions of the discrete modulated signals should be close to Gaussian distributions, especially for high rate communication systems or high order modulation communication systems. For example, the capacity-achieving superposition coded modulation (SCM)\cite{Wachsmann1999, Gadkari1999}, the quantization and mapping method, Gallager mapping \cite{Cover2006,Gamal2012}, etc. are widely used to generate Gaussian-like transmit signals. As long as the discrete sources adopt some of these Gaussian-like modulation and coding schemes, the theorems and results in this paper are expected to be applicable. In the simulation results, we will show some capacity-achieving Bit Error Ratio (BER) performances for practical discrete communication systems in which the superposition coded modulation is used to produce Gaussian-like transmit signals.}

\subsection{Existing Algorithms}
In this section, we first review the existing MMSE detection and classical iterative algorithms that are usually used in MU-MIMO systems. Then, a modification will be made on these algorithms. Specifically, the output estimation will be modified by taking the source distributions into consideration under the message passing rules. The results obtained will be used for the convergence analysis of the GMPID.

\begin{figure*}[t]
  \centering
  \includegraphics[width=16cm]{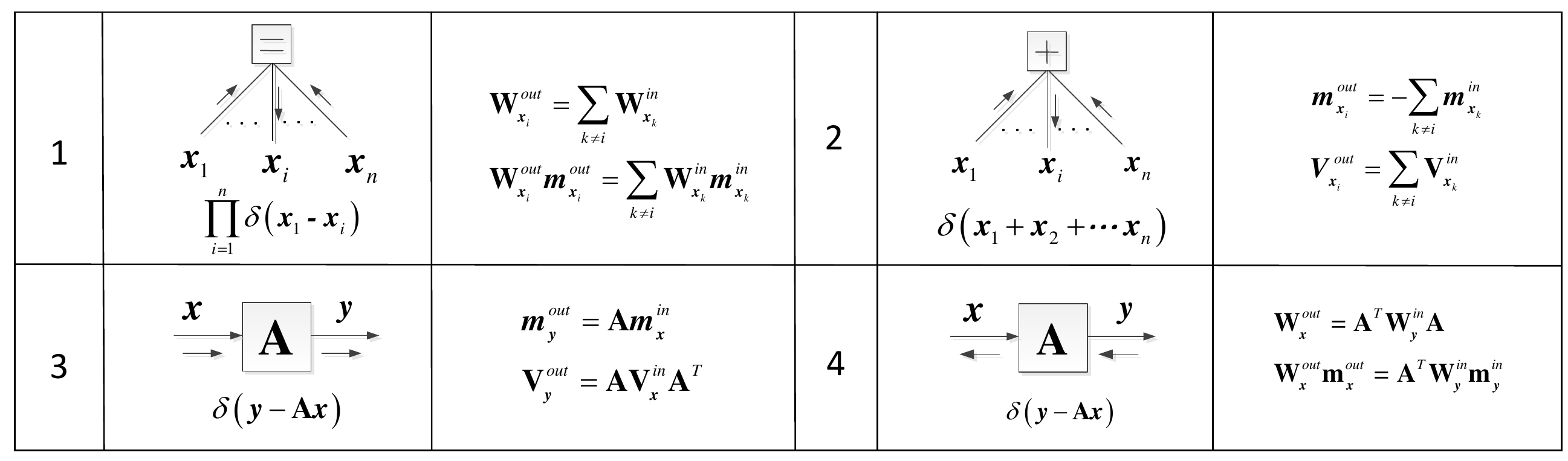}\\
  \caption{Message update rules for the equality constraint, sum constraint and matrix multiplication constraint.}\label{update_rule}
\end{figure*}
Some preliminaries about the message update rules \cite{kschischang2001, Loeliger2004, Loeliger2002, Loeliger2006, Guo2008} are given in Fig. \ref{update_rule}, which include the mean vector $\textbf{\emph{m}}$, the covariance matrix $\mathbf{V}$ and $\mathbf{W}\!\!=\!\!\mathbf{V}^{-1}$. If the messages are scalars, then the expressions can be replaced by scalar version.
\subsubsection{MMSE Detector}
It is well known that MMSE detection is optimal under MSE measure when the sources are Gaussian distributed \cite{verdu1998}. The MMSE detector \cite{tse2005} is given by
\begin{equation}\label{mmse5}
\tilde{x}_k=\frac{{{\bf{h}}_k^T\mathbf{V}_{{{\widetilde {\textbf{\emph{n}}}}_k}}^{ - 1}}\textbf{\emph{y}}}{{{\bf{h}}_k^T\mathbf{V}_{{{\widetilde {\textbf{\emph{n}}}}_k}}^{ - 1}{{\bf{h}}_k}}}=x_k + {n'_k},
\end{equation}
where $\mathbf{h}_i$ denotes the $i$th column of channel matrix $\mathbf{H}$, and  ${\mathbf{V}_{\widetilde {\textbf{\emph{n}}}_k}} = \sigma _n^2{\mathbf{I}_{\rm{\textbf{\emph{n}}}}} + \sum\limits_{i \ne k} {\sigma _{{x_i}}^2{\mathbf{h}_i}\mathbf{h}_i^T}$ denotes the covariance matrix of interference-noise vector $\sum\limits_{i\neq k} {{\mathbf{h}_i}{x_i} + \textbf{\emph{n}}}$.
The equivalent Gaussian noise satisfies ${n'_k}\sim \mathcal{N}(0,({{{\bf{h}}_k^T\mathbf{V}_{{{\widetilde {\textbf{\emph{n}}}}_k}}^{ - 1}{{\bf{h}}_k}}})^{-1})$. According to the first equality constraint update rule in Fig. \ref{update_rule}, each user combines the estimated distribution $\mathcal{N}(\tilde {x}_k,\sigma_{n'_k}^2)$ with the source distribution $\mathcal{N}(0,\sigma_{x_k}^2)$. We then get the following modified MMSE detector with the \emph{Matrix Inversion Lemma}.
\begin{equation}\label{mmse_m}
{{\hat {\textbf{\emph{x}}}}} = \sigma _{{{n}}}^{ - 2}\mathbf{V} _{{{\hat {\textbf{\emph{x}}}}}}\mathbf{H}^T\textbf{\emph{y}},
\end{equation}
where
\begin{equation}
\begin{array}{l}\label{mmse_var}
{\mathbf{V}_{\hat {\textbf{\emph{x}}}}} = {(\sigma _n^{ - 2}{\mathbf{H}^T}\mathbf{H} + \mathbf{V}_{\textbf{\emph{x}}}^{ - 1})^{ - 1}}\\
\quad  = {\mathbf{V}_\textbf{\emph{x}}} - {\mathbf{V}_\textbf{\emph{x}}}{\mathbf{H}^T}{({\sigma _n^2}{\mathbf{I}_M} + \mathbf{H}{\mathbf{V}_\textbf{\emph{x}}}{\mathbf{H}^T})^{ - 1}}\mathbf{H}{\mathbf{V}_\textbf{\emph{x}}}.
\end{array}
\end{equation}
The ${\mathbf{V}_{\hat {\textbf{\emph{x}}}}}$ contains the estimation error of each source. Specifically, the $k$th diagonal element $v_{kk}$ of the covariance matrix $\mathbf{V}_{\hat {\textbf{\emph{x}}}}$ denotes the estimation error of the source $x_k$.

\subsubsection{Classical Iterative Algorithms}
We express the iterative algorithms \cite{Axelsson1994} in a simple form
\begin{equation}\label{a1}
{\textbf{\emph{x}}}(t) = \mathbf{B}{\textbf{\emph{x}}}(t - 1) + {\textbf{\emph{c}}},
\end{equation}
where neither the iteration matrix $\mathbf{B}$ nor the vector ${\textbf{\emph{c}}}$ depends upon the iteration number $t$. Some iterative detections like \emph{Jacobi iterative detector} and \emph{Richardson iterative algorithm} are special cases of the classical iterative algorithms (\ref{a1}).

\textbf{\textit{Proposition 1 }}\cite{Bertsekas1989,Henrici1964}: \emph{Assuming that the matrix $\mathbf{I}-\mathbf{B}$ is invertible, the iteration (\ref{a1}) converges to the exact solution ${\textbf{{x}}}^*=(\mathbf{I}-\mathbf{B})^{-1}\mathbf{c}$ for any initial guess ${\textbf{{x}}}\left(0\right)$ if $\mathbf{I}-\mathbf{B}$ is strictly (or irreducibly) diagonally dominant or $\rho \left( \mathbf{B }\right) < 1$, where $\rho(\mathbf{B})$ is the spectral radius of $\mathbf{B}$.}

This convergence proposition of the classical iterative algorithms (\ref{a1}) is very important for the convergence analysis of the GMPID algorithm in following sections.

\subsection{Performance Analysis of the MMSE Detector}
Firstly, we introduce some results in \emph{Random Matrix Theory} that will be used in this subsection. When $\beta= K/M$ is fixed and $K\to \infty$, we can have the following expression \cite{verdu1998}.
\begin{equation}\label{PA1}
\frac{1}{K}{\rm{tr}}\left\{ {{{(\mathbf{I}_K + \eta {\mathbf{H}^T}\mathbf{H})}^{ - 1}}} \right\} \to 1 - \frac{{\mathcal{F}({\eta M },\beta )}}{{{{4\eta \beta M} }}}
\end{equation}
where $\eta$ is a constant and
\begin{equation}\label{PA2}
\!\mathcal{F}(x,z) \!=\! \!{\left(\!\! {\sqrt {x{{\left( {1 + \sqrt z } \right)}^2} + 1}  - \sqrt {x{{\left( {1 - \sqrt z } \right)}^2} + 1} } \right)^2}.
\end{equation}

From (\ref{mmse_var}), the MSE of the MMSE detector is calculated by
\begin{equation}\label{PA5}
MSE=\frac{1}{K}{\rm{tr}}(\mathbf{V}_{\hat {\textbf{\emph{x}}}}) = \frac{1}{K}{\rm{tr}}\{(\sigma _{{{n}}}^{ - 2}\mathbf{H}^T\mathbf{H}+\mathbf{V} _{{{ {\textbf{\emph{x}}}}}}^{-1})^{-1}\}.
\end{equation}
Assuming that $\mathbf{V}_{\textbf{\emph{x}}}=\sigma_x^2\mathbf{I}_K$ and with (\ref{PA1}), (\ref{PA2}) and (\ref{PA5}), we can get the following proposition.

\textbf{\textit{Proposition 2:}} \emph{For the Massive MU-MIMO system where $\beta\!\!= \!\!K/M$ is fixed, $K\!\!\to\!\! \infty$, and the transmitted symbols of the sources are i.i.d. with $\mathcal{N}^K(0,\sigma_x^2)$, the MSE of the optimal MMSE detector is described by
\begin{equation}\label{PA6}
\!\!\!\!\!\!\!\!MSE \to  {\sigma _x^2 \!-\! \frac{\sigma _x^2}{{4snr\beta M}}{\cal F}(snrM,\beta )}  \to \!\!\left\{ \begin{array}{l}
\!\!\!\frac{{\sigma _n^2}}{{M - K}}\mathop {}\limits_{\mathop {}\limits_{} } ,\;\beta  < 1\\
\!\!\!\frac{{K - M}}{K}\sigma _x^2\mathop {}\limits_{\mathop {}\limits_{} } ,\beta  > 1\\
\!\!\!\sqrt {\frac{{\sigma _x^2\sigma _n^2}}{K}} ,\; \beta  = 1.
\end{array} \right.
\end{equation}
where $snr={{\sigma _x^2} \mathord{\left/
 {\vphantom {{\sigma _x^2} {\sigma _n^2}}} \right.
 \kern-\nulldelimiterspace} {\sigma _n^2}}$ is the signal-to-noise ratio.}

We can obtain some interesting results from \emph{Proposition 2}. Firstly, when $K<M$, the MSE of MMSE detection is determined by the variance of the Gaussian noise and the value of $M-K$, but it is independent of the variances of the sources. Secondly, when $M<K$, the MSE of MMSE detection is determined by the variances of the sources with a scale parameter $\frac{{{\rm{K}} - M}}{K}$, but it is independent of the Gaussian noise. This means that the system keeps the same MSE even if we decrease the variance of the noise at the BS. Finally, when $M=K$, the MSE of MMSE detection depends on the variance of the Gaussian noise, the variances of the sources and the number of users $K$. From (\ref{PA6}), we can see that the performance of the massive MU-MIMO system is poor when $\beta\geq1$. Hence, we only consider the case $\beta<1$ in this paper.

\textbf{\textit{Remark 1}}: {Although the result in (\ref{PA6}) is derived for the asymptotic MSE analysis of MU-MIMO MMSE detection, it also comes close to the actual MSE of the MU-MIMO system with small values of $K$ and $M$ \cite{verdu1998}.

\subsection{Complexity of MMSE Detector}
{From (\ref{mmse_m}) and (\ref{mmse_var}), we can see that the complexity of MMSE detection is $\mathcal{O}(\min\{M^3+KM^2,K^3+MK^2\})$, where $\mathcal{O}(K^3)$ (or $\mathcal{O}(M^3)$) arises from the matrix inverse calculation and $\mathcal{O}(MK^2)$ (or $\mathcal{O}(KM^2)$) arises from the matrix multiplication. The complexity of MMSE detection is very high when the number of users and the number of antennas are very large. Therefore, the research on low-complexity detectors without performance loss for the massive MU-MIMO systems is important. In the next section, we consider a low-complexity \emph{Gaussian Message Passing Iterative Detector} for the massive MU-MIMO systems, which can converge to the optimal MMSE detector.}

\section{Gaussian Message Passing Iterative Detector}
The GMPID for the MU-MIMO systems is based on a pairwise factor graph, as shown in Fig. \ref{f2}. Similar to the Belief Propagation (BP) decoding process of LDPC code \cite{William2009}, the GMPID calculates the output message, called extrinsic information, on each edge by employing the messages on the other edges that are connected with the same node. There are two main differences between the GMPID and the BP decoding process of LDPC code, one of which is that the messages passed on each edge of GMPID are the means and variances, while the BP decoding process passes the likelihood values. The second difference is the different message update functions at the sum nodes and variable nodes. Fig. \ref{f3} presents the message updating diagram of the GMPID. The message updating rules are given as follows.

\begin{figure}[t]
  \centering
  \includegraphics[width=9cm]{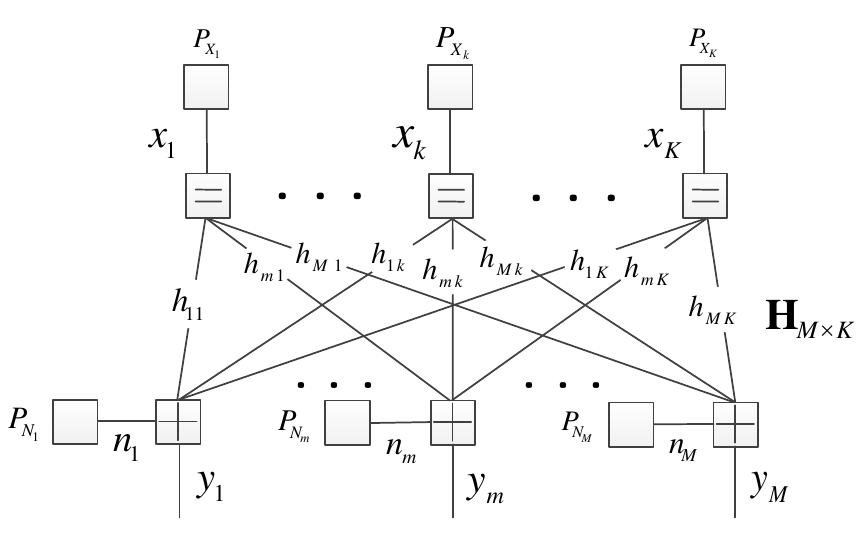}\\
  \caption{Gaussian message passing iterative Detection for MU-MIMO systems.}\label{f2}
\end{figure}
\begin{figure*}[t]
  \centering
  \includegraphics[width=14cm]{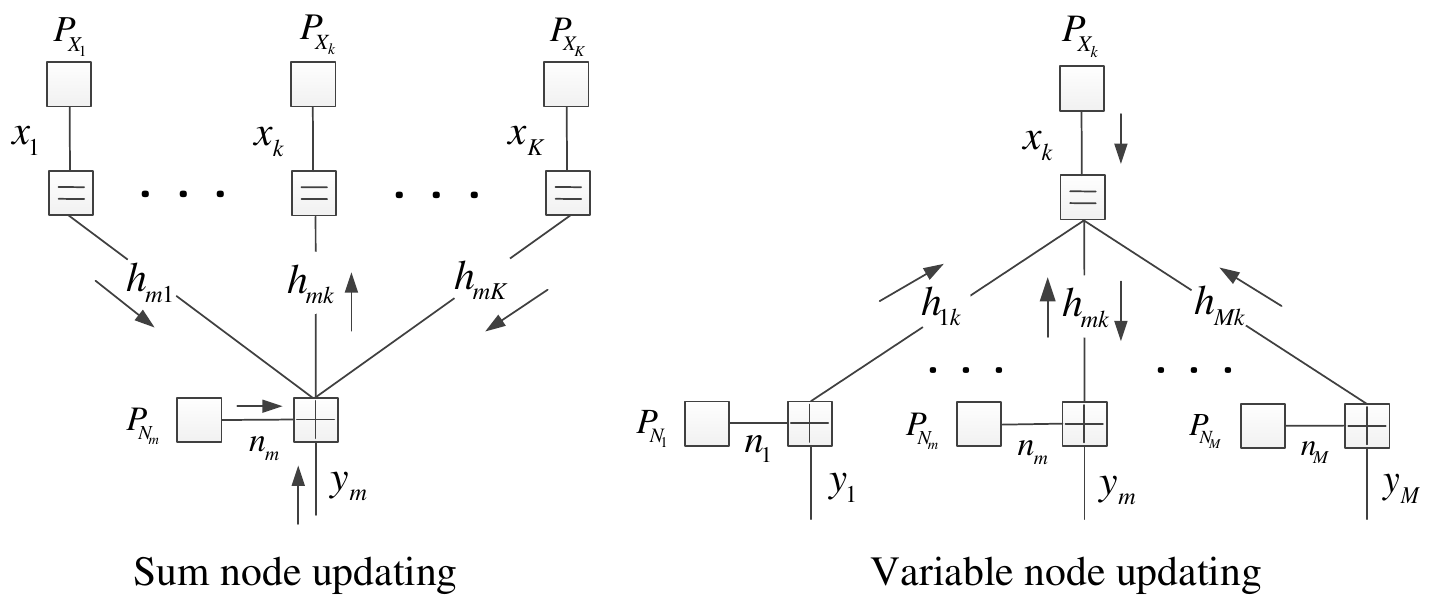}\\
  \caption{Message update at sum nodes and variable nodes. The output message called extrinsic information on each edge is calculated by the messages on the other edges that are connected with the same node. The messages passed on each edge are the mean and variance of a Gaussian distribution.}\label{f3}
\end{figure*}
\subsection{Message Update at Sum Nodes of GMPID}
Each sum node can be seen as a multiple-access process and its message is updated by
\begin{equation}\label{np1}
\left\{ \begin{array}{l}
e_{m \to k}^s(t) = {y_m} - \sum\limits_{i \ne k} {h_{mi}e_{i \to m}^v(t-1)\;,} \\
v_{m \to k}^s(t)= \sum\limits_{i \ne k} {h_{mi}^2v_{i \to m}^v(t-1) + \sigma _n^2\;},
\end{array} \right.
\end{equation}
where $i,k \in \{ 1,2, \cdots ,K\} ,m \in \{ 1,2, \cdots ,M\}$, $y_m$ is the $m$th element of the received vector $\textbf{\emph{y}}$, $h_{mi}$ is the element of channel matrix $\mathbf{H}$ in $m$th row and $i$th column, and $\sigma^2_n$ is the variance of the Gaussian noise. In addition, $e_{k \to m}^v(t)$ and $v_{k \to m}^v(t)$ denote the mean and variance passing from the $k$th variable node to $m$th sum node respectively, and $e_{m \to k}^s(t)$ and $v_{m \to k}^s(t)$ denote the mean and variance passing from $m$th sum node to $k$th variable node respectively. The initial value $\mathbf{v}^v(0)$ equals to $+\boldsymbol{\infty}$ and $\mathbf{e}^v(0)$ equals to $\mathbf{0}$, where $\mathbf{v}^v(t)$ and $\mathbf{e}^v(t)$ are vectors containing the elements $v_{k\to m}^v(t)$ and $e_{k\to m}^v(t)$ for all $k \in \{ 1,2, \cdots ,K\}$ and $m \in \{ 1,2, \cdots ,M\}$ respectively.

\subsection{Message Update at Variable Nodes of GMPID}
Each variable node can be seen as a broadcast process and its message update is denoted by
\begin{equation}\label{var_mes}
\left\{ {\begin{array}{*{20}{l}}
{v_{k \to m}^v(t) = {{( {\sum\limits_{i} {h_{ik}^2v_{i \to k}^{s{\;^{ - 1}}}(t) + \sigma _{{x_k}}^{ - 2}\;} } )}^{ - 1}},}\\
{e_{k \to m}^v(t) = v_{k \to m}^v(t)\sum\limits_{i } {{h_{ik}}v_{i \to k}^{s{\;^{ - 1}}}(t)e_{i \to k}^s(t)\;\;} }.
\end{array}} \right.\quad
\end{equation}
where $k \in \{ 1,2, \cdots ,K\}, i,m \in \{ 1,2, \cdots ,M\}$, and $\sigma^2_{x_k}$ denotes the variance of the source $x_k$.

\subsection{Decision and Output of GMPID}
When the MSE of the GMPID meets the requirement or the number of iterations reaches the limit, we output the estimation $\hat{x}_{k}$ and its MSE $\sigma^2_{\hat{x}_{k}}$ of $x_k$ as follows.
\begin{equation}\label{np2}
\left\{ \begin{array}{l}
{\sigma}^2_{\hat{x}_k} = {( {\sum\limits_{m } {h_{mk}^2v_{m \to k}^{s{\;^{ - 1}}}(t) + \sigma _{{x_k}}^{ - 2}\;} } )^{ - 1}},\\
\hat{x}_k = {\sigma}^2_{\hat{x}_k}\sum\limits_{m } {h_{mk}v_{m \to k}^{s{\;^{ - 1}}}(t)e_{m \to k}^s(t)\;\;},
\end{array} \right.\quad
\end{equation}
where $k \in \{ 1,2, \cdots ,K\}$. It should be pointed out that the decision is made based on the full information coming from all the sum nodes.

{\textbf{\emph{Remark 2}}: Compared with the original GMPID, there is a minor modification in the variable node updating rule in (\ref{var_mes}). The modification is that the full information, not just the extrinsic information, is passed from the variable nodes to the sum nodes. This does not lead to significant performance loss, because the difference between these two messages is negligible at the variable nodes when $M$ is very large and $K<M$. In fact, we can show that the variances of the GMPID converge to those of the original GMPID, which means that both algorithms have the same performance. Our simulation results will also show that the modified GMPID and the original GMPID have the same performance when used in the massive MU-MIMO detection.

The challenge of the original GMPID is that it is hard to analyse its mean convergence directly because the structure of the original GMPID algorithm is too complicated. Interestingly, when using the message update (\ref{var_mes}) at the variable node, the GMPID can be rewritten into a matrix form like the classical iterative algorithm. Thus, the convergence analysis of the GMPID becomes feasible. Based on the above considerations, in the rest of this paper, we do not distinguish the modified GMPID with the original one and will call them GMPID. In the next section, we will give the convergence analysis of the variances and the means of GMPID respectively.}

\subsection{GMPID in Matrix Form}
Let $\mathbf{A}_{M\times N}.*\mathbf{B}_{M\times N}=\left[ a_{ij}b_{ij}\right]_{M\times N}$, $diag^{-1}\{\mathbf{A}_{N\times N}\}=\left[a_{11},a_{22},\cdots,a_{NN}\right]^T$, $\mathbf{1}_{M\!\times \!N}\!\!=\!\!\left[1\right]_{M\!\times \!N}$ and $\mathbf{A}^{(k)}_{M\!\times \!N}\!\!=\!\!\left[a^k_{ij}\right]_{M\!\times \!N}$. Assume $\mathbf{E}_{su}(t)=\left[e_{m\to i}^s(t)\right]_{M\times K}$, $\mathbf{V}_{su}(t)=\left[v_{m\to i}^s(t)\right]_{M\times K}$, $\mathbf{E}_{us}(t)=\left[e_{i\to m}^v(t)\right]_{K\times M}$ and $\mathbf{V}_{us}(t)=\left[v_{i\to m}^v(t)\right]_{K\times M}$.

The message update at the sum nodes (\ref{np1}) is rewritten as
\begin{equation}\label{e5}
\begin{array}{l}
\!\!\!\left[\!\!\! \begin{array}{l}
{{\rm{\mathbf{E}}}_{su}}(t)\\
{\mathbf{V}_{su}}(t)
\end{array} \!\!\!\right]\!\! =\!\! \left[ \!\!\!\begin{array}{l}
\quad\;\;\textbf{\emph{y}} - dia{g^{ - 1}}\{ \mathbf{1}_{M\times K}\cdot{\widetilde{\mathbf{E}}_{us}}(t - 1)\} \\
\sigma_n^2\cdot{\mathbf{1}}_{M\times1}\!+\!dia{g^{ - 1}}\{ \mathbf{1}_{M\times K}\!\cdot \!{\widetilde{\mathbf{V}}_{us}}(t \!-\! 1)\}
\end{array} \!\!\!\right] \!\cdot\! {\mathbf{1}_{1 \times K}}\mathop {}\limits_{\mathop {}\limits_{\mathop {}\limits_{} } } \\
\qquad\qquad\;\;-\left[ \begin{array}{l}
-\widetilde{\mathbf{E}}_{us}^T(t - 1)\\
\widetilde{\mathbf{V}}_{us}^T(t - 1)
\end{array} \right],
\end{array}
\end{equation}
where $\widetilde{\mathbf{E}}_{us}(t)=\mathbf{E}_{us}(t).*\mathbf{H}^T$ and $\widetilde{\mathbf{V}}_{us}(t)=\mathbf{V}_{us}(t).*{\mathbf{H}^{(2)}}^{T}$.
Let $\mathbf{v}_{\textbf{\emph{x}}}=[\sigma_{x_1}^2,\ldots, \sigma_{x_1}^2]^T$, $\mathbf{W}_{su}(t)=\mathbf{V}_{su}^{(-1)}(t)$, $\mathbf{W}_{us}(t)={\mathbf{V}_{us}^{(-1)}(t)}$ and $\mathbf{G}_{us}(t)=\mathbf{W}_{us}(t).*\mathbf{E}_{us}(t)$. At the variable nodes, the message update (\ref{var_mes}) is rewritten as
\begin{equation}\label{e6}
\begin{array}{l}
\left[ \!\!\begin{array}{l}
{\mathbf{W}_{us}}(t)\\
{\mathbf{G}_{us}}(t)
\end{array} \!\!\right] \!=\! \left[ \!\!\begin{array}{l}
\mathbf{v}_{\textbf{\emph{x}}}^{(-1)} \!+\! dia{g^{ - 1}}\left\{ {{\mathbf{1}_{K \times M}} \cdot {{\widetilde{\mathbf{W}}_{su}}(t)}} \right\}\mathop {}\limits_{\mathop {}\limits_{} }\\
\quad\;\; dia{g^{ - 1}}\left\{ {{\mathbf{1}_{K \times M}} \cdot \widetilde{\mathbf{G}}_{su}(t)} \right\}
\end{array}\!\!\! \right]\! \cdot\! {\mathbf{1}_{1 \times M}}
,
\end{array}
\end{equation}
where $\widetilde{\mathbf{W}}_{su}(t)=\mathbf{H}^{(2)}.*\mathbf{W}_{su}(t)$ and $\widetilde{\mathbf{G}}_{su}(t)=\mathbf{H}.*\mathbf{G}_{su}(t)=\mathbf{H}.*\mathbf{W}_{su}(t).*\mathbf{E}_{su}(t)$.
 Then, we can get $\mathbf{V}_{us}(t)$ by $\mathbf{V}_{us}(t)=\mathbf{W}_{us}^{(-1)}(t)$, and get $\mathbf{E}_{us}(t)$ by $\mathbf{E}_{us}(t)=\mathbf{V}_{us}(t).*\mathbf{G}_{us}(t)$.

 When$|\mathbf{E}_{us}(t)-\mathbf{E}_{us}{(t-1)}|<\epsilon$, where $\epsilon$ is a sufficiently small positive precision parameter, or when the evaluation has passed a sufficient number of iterations, we output $\hat{\textbf{\emph{x}}}$ as the estimation of $\textbf{\emph{x}}$ and output the MSE $\mathbf{\sigma}_{\hat{\textbf{\emph{x}}}}^2$ between $\textbf{\emph{x}}$ and $\hat{\textbf{\emph{x}}}$ by
\begin{equation}\label{e7}
\left\{ \begin{array}{l}
{\mathbf{\sigma}^2_{\hat{\textbf{\emph{x}}}}} = {\left( {dia{g^{ - 1}}\left\{ {{\mathbf{1}_{K \times M}} \cdot {\widetilde{\mathbf{W}}_{su}}(t) } \right\}} \right)^{( - 1)}}\!\!\!\!,\\
{\hat{\textbf{\emph{x}}}} = {{\mathbf{\sigma}^2_{\hat{\textbf{\emph{x}}}}}}.*dia{g^{ - 1}}\left\{ {\mathbf{1}_{K \times M}} \cdot \widetilde{\mathbf{G}}_{su}(t) \right\}.
\end{array} \right.
\end{equation}

\begin{algorithm}[t]
\caption{GMPID Algorithm}
\begin{algorithmic}[1]
\State {\small{\textbf{Input:} {{$\mathbf{H}$, $\mathbf{V}_{\textbf{\emph{x}}}$, $\sigma^2_n$, $\epsilon>0$, $N_{ite}$}} and calculate {{$\mathbf{H}^{(2)}$}}.
\State \textbf{Initialization:} {{$t\!=\!-1$, $\mathbf{E}_{us}(0)\!\!=\!\!\left[0\right]_{K\times M}$}}, {{$\mathbf{V}_{us}(0)\!=\!\left[+\infty\right]_{K\times M}$.}}
\State \textbf{Do}
\State { \;\;$t\!=\!t\!+\!1$, $\!\widetilde{\mathbf{E}}_{us}(t) \!=\! \mathbf{E}_{us}(t).*\!\mathbf{H}^T\!$ and $\!\widetilde{\mathbf{V}}_{us}(t)\!=\!\mathbf{V}_{us}(t).*{\mathbf{H}^{(2)}}^{T}$\!\!\!\!,
\State  \vspace{-0.25cm}\[\begin{array}{l}\left[ \!\!\!\begin{array}{l}
{{\rm{\mathbf{E}}}_{su}}(t)\\
{\mathbf{V}_{su}}(t)
\end{array} \!\!\!\!\right] \!\!=\!\! \left[ \!\!\!\begin{array}{l}
\quad\;\;\textbf{\emph{y}} - dia{g^{ - 1}}\{ \mathbf{1}_{M\times K}\cdot{\widetilde{\mathbf{E}}_{us}}(t - 1)\} \\
\sigma_n^2\!\cdot\!{\mathbf{1}}_{M\times1}\!+\!dia{g^{ - 1}}\{ \mathbf{1}_{M\times K}\!\cdot\! {\widetilde{\mathbf{V}}_{us}}(t - 1)\}
\end{array} \!\!\!\!\right] \!\!\cdot\!\! {\mathbf{1}_{1 \times K}}\mathop {}\limits_{\mathop {}\limits_{\mathop {}\limits_{} } }  \\
\qquad\quad\qquad-\left[ \begin{array}{l}
-\widetilde{\mathbf{E}}_{us}^T(t - 1)\\
\widetilde{\mathbf{V}}_{us}^T(t - 1)
\end{array} \right],\qquad\qquad\qquad\end{array}\]}}
\State\vspace{-0.35cm}\[\;\; \widetilde{\mathbf{W}}_{su}(t)\!=\!\mathbf{H}^{(2)}.*\mathbf{V}^{(-1)}_{su}(t), \widetilde{\mathbf{G}}_{su}(t)\!=\!\mathbf{H}.*\mathbf{V}^{(-1)}_{su}(t).*\mathbf{E}_{su}(t),\]
\State \vspace{-0.35cm}\[ \begin{array}{l}
\left[ \!\!\begin{array}{l}
{\mathbf{W}_{us}}(t)\\
{\mathbf{G}_{us}}(t)
\end{array} \!\!\!\right] \!\!=\!\! \left[ \!\!\begin{array}{l}
\mathbf{v}_{\textbf{\emph{x}}}^{(-1)} \!+\! dia{g^{ - 1}}\{ {{\mathbf{1}_{K \times M}} \!\cdot\! {{\widetilde{\mathbf{W}}_{su}}(t)}}\}\mathop {}\limits_{\mathop {}\limits_{} }\\
\quad\;\; dia{g^{ - 1}}\{ {{\mathbf{1}_{K \times M}} \cdot \widetilde{\mathbf{G}}_{su}(t)}\}
\end{array} \!\!\!\right] \cdot {\mathbf{1}_{1 \times M}},
\end{array}\qquad\qquad\qquad\]
\vspace{0cm}
\State \; $\mathbf{V}_{us}(t)=\mathbf{W}_{us}^{(-1)}(t)$ and  $\mathbf{E}_{us}(t)=\mathbf{V}_{us}(t).*\mathbf{G}_{us}(t)$.
\vspace{0.15cm}
\State \textbf{While} \;{\small{$\left( \:|\mathbf{E}_{us}(t)-\mathbf{E}_{us}{(t-1)}|<\epsilon \;{\textbf{or}}\; t\leq N_{ite} \;\right)$}}
\State  \vspace{-0.25cm}{{ \[ \begin{array}{l}{\mathbf{\sigma}^2_{\hat{\textbf{\emph{x}}}}} = {\left( {dia{g^{ - 1}}\{ {{\mathbf{1}_{K \times M}} \cdot {\widetilde{\mathbf{W}}_{su}}(t) }\}} \right)^{( - 1)}}\!\!\!\!,\\
{\hat{\textbf{\emph{x}}}} = {{\mathbf{\sigma}^2_{\hat{\textbf{\emph{x}}}}}}.*dia{g^{ - 1}}\{ {\mathbf{1}_{K \times M}} \cdot \widetilde{\mathbf{G}}_{su}(t) \}.\end{array} \qquad\qquad\qquad\qquad\]}}
\State \textbf{Output:}  ${\hat{\textbf{\emph{x}}}}$ and ${\mathbf{\sigma}^2_{\hat{\textbf{\emph{x}}}}}$ }.
\end{algorithmic}
\end{algorithm}

\subsection{Complexity of GMPID}
This section will show that the matrix GMPID form further reduces the complexity and also permits a parallel iterative detection algorithm. As the variance calculations are independent of the received signals $\textbf{\emph{y}}$ and the means, it can be pre-computed before the iteration. In each iteration, it needs about $4KM$ multiplications and $4KM$ additions. Therefore, the complexity is as low as $\mathcal{O}(KMN_{ite})$, where $N_{ite}$ is the number of iterations. The distributed scalar operation at each node of the GMPID avoids the huge matrix calculation, which results in a lower complexity. Algorithm 1 shows the detailed process of the GMPID.

\subsection{Variance Convergence of GMPID}
 We show the variance convergence of GMPID by the following proposition.

\textbf{\textit{Proposition 3:}} \emph{For the Massive MU-MIMO system where $\beta\!=\! K/M$ is fixed, $K\!\to\! \infty$, and the transmitted symbols of the sources are i.i.d. with $\mathcal{N}^K\!(0,\sigma_x^2)$, the variances of GMPID converge to\vspace{-0.2cm}
\begin{equation}\label{p6}
\sigma _{\hat x}^2\approx {\hat \sigma ^2}\approx \left\{ \begin{array}{l}
\frac{{\sigma _n^2}}{{M - K + sn{r^{ - 1}}}}\mathop {}\limits_{\mathop {}\limits_{} } ,\quad \beta  < 1,\\
\frac{{K - M }}{K}\sigma _x^2\mathop {}\limits_{\mathop {}\limits_{} } ,\quad \beta  > 1,\\
\sqrt {\frac{{\sigma _x^2\sigma _n^2}}{K}} ,\quad \beta  = 1,
\end{array} \right.
\end{equation}
where $snr={{\sigma _x^2} \mathord{\left/
 {\vphantom {{\sigma _x^2} {\sigma _n^2}}} \right.
 \kern-\nulldelimiterspace} {\sigma _n^2}}$ is the signal-to-noise ratio.}

\begin{IEEEproof}
From (\ref{np1}) and (\ref{var_mes}), we have
\begin{equation}\label{p1}
\!\!{v_{k \to m}^v(t) \!=\!\! {{\left( \!{\sum\limits_{i} {h_{ik}^2{{( {\sum\limits_{j \ne k} {h_{ij}^2v_{j \to i}^v(t\!-\!1) \!+\! \sigma _n^2\;} } )}^{ - 1}} \!\!\!\!\!\!\!+ \sigma _{{x_k}}^{ - 2}\;} } \!\!\right)}^{ - 1}}}\!\!\!\!\!\!.
\end{equation}
As the initial value $\mathbf{v}^v(0)$ is equal to $+\boldsymbol{\infty}$, it is easy to see that $\mathbf{v}^v(t) > 0$ for any $t>0$ during the iteration. Hence, $\mathbf{v}^v(t)$ has a lower bound $\mathbf{0}$. From (\ref{p1}), we can see that $\mathbf{v}^v(t)$ is a monotonically non-increasing function with respect to $\mathbf{v}^v(t-1)$. Besides, we can get $\mathbf{v}^v(1)<\mathbf{v}^v(0)=+\boldsymbol{\infty}$ for the first iteration. All the inequations in this paper are component-wise inequalities. Therefore, it can be shown that $\mathbf{v}^v(t)\leq \mathbf{v}^v(t-1)$ with $\mathbf{v}^v(1) \leq \mathbf{v}^v(0)$ by the monotonicity of the iteration function. This means that $\{\mathbf{v}^v(t)\}$ is a monotonically decreasing sequence but is lower bounded. Thus, sequence $\{\mathbf{v}^v(t)\}$ converges to a certain value, i.e., $\mathop {\lim }\limits_{t \to  \infty } {\mathbf{v}^v}(t)=\mathbf{v}^*$.

To simplify the calculation, we assume $\mathbf{V}_{\textbf{\emph{x}}}=\sigma_x^2\mathbf{I}_K$, i.e., $\sigma_{x_k}^2=\sigma_x^2$ for $k \in \{ 1, \cdots ,K\}$. With the symmetry of all the elements in $\mathbf{v}^*$, we can get $v_{k\to m}^*=\hat{\sigma}^2$ for $k \in \{ 1, \cdots ,K\}$ and $m \in \{ 1, \cdots ,M\}$. Thus, from (\ref{p1}), the convergence point $\hat{\sigma}^2$ can be solved by
\begin{equation}\label{p2}
{\hat{\sigma}^2 = {{\left( {\sum\limits_{i } {h_{ik}^2{{(\hat{\sigma}^2 {\sum\limits_{j \ne k} {h_{ij}^2 + \sigma _n^2\;} } )}^{ - 1}} + \sigma _{{x}}^{ - 2}\;} } \right)}^{ - 1}}}.
\end{equation}
As the channel parameters $h_{ik}^2$ and $h_{ij}^2$ are independent with each other, the above expression can be rewritten as
\begin{equation}\label{p3}
\sigma _{{x}}^{ - 2}\!\sum\limits_{j \ne k} {h_{ij}^2} {{\hat \sigma }^4} \!+\! (\sigma _n^2\sigma _{{x}}^{ - 2} \!+\!\! \sum\limits_{i} {h_{ik}^2- \sum\limits_{j \ne k} {h_{ij}^2\;} } ){{\hat \sigma }^2} - \sigma _n^2 = 0.
\end{equation}
When $M$ is large, taking an expectation on (\ref{p3}) with respect to the channel parameters $h_{ik}^2$ and $h_{ij}^2$, we get
\begin{equation}\label{p4}
K\sigma _{{x}}^{ - 2}{{\hat \sigma }^4} + (\sigma _n^2\sigma _{{x}}^{ - 2} + M - K ){{\hat \sigma }^2} - \sigma _n^2 = 0.
\end{equation}
Then $\hat{\sigma}^2$ is the positive solution of (\ref{p4}), i.e.,
\begin{equation}\label{p42}
\!\!{{\hat \sigma }^2}\!\! =\!\! \frac{{\sqrt {{{\!(\sigma _n^2\sigma _x^{ - 2} \!+ \!M\! -\! K)}^2}\!\!\! +\! 4K \sigma _x^{ - 2}\sigma _n^2} \! - \!(\sigma _n^2\sigma _x^{ - 2} \!+ \!M\! - \!K)}}{{2K \sigma _x^{ - 2}}}.
\end{equation}
With (\ref{np2}) and(\ref{p42}),\emph{ Proposition 3} is proved.
\end{IEEEproof}

As $snr^{-1}$ is an infinitesimal compared with $K$ and $M$ and thus can be ignored, we can see that (\ref{p6}) has the same MSE performance as the MMSE detection given in (\ref{PA6}). Thus, we obtain the following theorem.

\textbf{\textit{Theorem 1:}} \emph{For the Massive MU-MIMO system where $\beta= K/M$ is fixed, $K\to \infty$, and the transmitted symbols of the sources are i.i.d. with $\mathcal{N}^K(0,\sigma_x^2)$, the variances of GMPID converge to the exact MSE of the MMSE detector.}

\textbf{\emph{Remark 3}}: Actually, the variance convergence analysis of GMPID is the same as that of the original GMPID. It is easy to find that the original GMPID has the same results on the variance convergence. Moreover, it should be pointed out that the above analysis provides an alternative way to estimate the MSE performance of the MMSE detector.

Similar to $\{\mathbf{v}^v(t)\}$, sequence $\{\mathbf{v}^s(t)\}$ also converges to a certain value, i.e., $v^s_{m\to k} \to \tilde{\sigma}^2$ for $k \in \{ 1, \cdots ,K\}$ and $m \in \{ 1, \cdots ,M\}$. From (\ref{np1}), we can get
\begin{equation}\label{p8}
{\tilde \sigma ^2} \approx K{\hat \sigma ^2} + \sigma _n^2.
\end{equation}
Let $\gamma={{{{\hat \sigma }^2}} \mathord{\left/
 {\vphantom {{{{\hat \sigma }^2}} {{{\tilde \sigma }^2}}}} \right.
 \kern-\nulldelimiterspace} {{{\tilde \sigma }^2}}}$, from (\ref{p6}) and (\ref{p8}), we get
 \begin{equation}\label{p7}
 \!\!\!\gamma  = \frac{1}{{{\rm{K}} + {{\sigma _n^2} \mathord{\left/
 {\vphantom {{\sigma _n^2} {{{\hat \sigma }^2}}}} \right.
 \kern-\nulldelimiterspace} {{{\hat \sigma }^2}}}}} \approx \left({M + sn{r^{ - 1}}} \right)^{ - 1}, \;\;\;\beta  < 1.
 \end{equation}

\subsection{Mean Convergence of GMPID}
Unlike the variances, the means of GMPID do not always converge. Two sufficient conditions for the mean convergence of GMPID are given by the following theorem.

\textbf{\textit{Theorem 2:}} \emph{For the Massive MU-MIMO system where $\beta= K/M$ is fixed, $K\to \infty$, and the transmitted symbols of the sources are i.i.d. with $\mathcal{N}^K(0,\sigma_x^2)$, the GMPID converges to the MMSE estimation if any of the following conditions holds.}

\emph{1. The matrix $\mathbf{I}_K + \gamma\left({\mathbf{H}^T}\mathbf{H}-\mathbf{D}_{{\mathbf{H}^T}\mathbf{H}}\right)$ is strictly or irreducibly diagonally dominant,}

\emph{2. $\rho \left( {\gamma ({\mathbf{H}^T}\mathbf{H} - {\mathbf{D}_{{\mathbf{H}^T}\mathbf{H}}})} \right) < 1$,}

\emph{\!\!\!\!\!\!where  $\gamma={{{{\hat \sigma }^2}} \mathord{\left/
 {\vphantom {{{{\hat \sigma }^2}} {{{\tilde \sigma }^2}}}} \right.
 \kern-\nulldelimiterspace} {{{\tilde \sigma }^2}}}$.}

\begin{IEEEproof}
From (\ref{np1}) and (\ref{var_mes}), we have
\begin{equation}\label{p9}
\!\!\!\!\!e_{k \to m}^v(t) \!\!=\! \!v_{k \to m}^v(t)\!\!\sum\limits_{i} {{h_{ik}}v_{i \to k}^{s{\;^{ - 1}}}(t)( {{y_i} \!-\!\!\! \sum\limits_{j \ne k} {{h_{ij}}e_{j \to i}^v(t \!-\! 1)} } )}.
\end{equation}
From the variance convergence analyses (\ref{p8}) and (\ref{p6}), the $v^s_{m\to k}$ and $v^v_{k\to m} $ converge to $\tilde{\sigma}^2$ and $\hat{\sigma}^2$ respectively. Therefore, (\ref{p9}) can be rewritten as
\begin{equation}\label{p10}
e_{k \to m}^v(t) = \frac{{\hat \sigma ^2}}{{\tilde \sigma ^2}}\sum\limits_{i} {{h_{ik}}( {\;{y_i} - \sum\limits_{j \ne k} {{h_{ij}}e_{j \to i}^v(t - 1)\;} } )\;}.
\end{equation}
Then, we can get $e_{k \to m}^v(t)=e^v_{k}(t)$, $k \in \{ 1, \cdots ,K\}$ and $m \in \{ 1, \cdots ,M\}$. Thus, the above equation is rewritten as
\begin{equation}\label{p11}
\mathbf{e}^v(t) =  \gamma{{\mathbf{H}^T}\textbf{\emph{y}} - \gamma\left({\mathbf{H}^T}\mathbf{H}-\mathbf{D}_{{\mathbf{H}^T}\mathbf{H}}\right) \mathbf{e}^v(t-1)},
\end{equation}
where $\gamma={{{{{\hat \sigma }^2}} \mathord{\left/
 {\vphantom {{{{\hat \sigma }^2}} {{{\tilde \sigma }^2}}}} \right.
 \kern-\nulldelimiterspace} {{{\tilde \sigma }^2}}}}$, $\mathbf{e}^v(t)={\left[ {{{e}_1^v}(t)\;{{e}_2^v}(t)\; \cdots \;{{e}_K^v}(t)} \right]^T}$, $\mathbf{D}_{\mathbf{H}^T\mathbf{H}}=diag\{d_{11},d_{22}, \cdots ,d_{KK}\}$ is a diagonal matrix and $d_{kk}=\mathbf{h}_k^T\mathbf{h}_k$, {\small$k\in \{1,2,\cdots,K\}$} is the diagonal element of the matrix $\mathbf{H}^T\mathbf{H}$. When $M$ is large, from the \emph{Law of Large Numbers}, the matrix $\mathbf{D}_{\mathbf{H}^T\mathbf{H}}$ can be approximated by $M\mathbf{I}_K$. Assuming that the sequence $\{\mathbf{e}^v(t)\}$ converges to $\mathbf{e}^*$, then from (\ref{p11}) we have
\begin{equation}\label{p12}
{\mathbf{e}^*} = {\left( {(\gamma^{-1} - M){\mathbf{I}_K} + {\mathbf{H}^T}\mathbf{H}} \right)^{ - 1}}{\mathbf{H}^T}\textbf{\emph{y}}.
\end{equation}
 When $\beta= K/M$ is fixed and $K\to \infty$, we have $\gamma^{-1} - M \to snr^{-1}$ from (\ref{p6}) and (\ref{p7}). With $\mathbf{V}_{\textbf{\emph{x}}}=\sigma_x^2\mathbf{I}_K$ and $\beta<1$, it is easy to see that it converges to the same value as (\ref{mmse_m}), i.e.,
\begin{equation}\label{p13}
{\mathbf{e}^*} = {\left( {snr^{-1}{\mathbf{I}_K} + {\mathbf{H}^T}\mathbf{H}} \right)^{ - 1}}{\mathbf{H}^T}\textbf{\emph{y}},
\end{equation}
which means that the GMPID converges to the MMSE estimation if it converges. Let $ \textbf{\emph{c}}= \gamma{\mathbf{H}^T}\textbf{\emph{y}}$ and $\mathbf{B}= \gamma\left({\mathbf{H}^T}\mathbf{H}-\mathbf{D}_{{\mathbf{H}^T}\mathbf{H}}\right)$, then (\ref{p11}) is a classical iterative algorithm (\ref{a1}). Thus, we can get \emph{Theorem 2} with \emph{Proposition 1}.
\end{IEEEproof}

As $\gamma \to \frac{1}{M+snr^{-1}}$, $K,M\rightarrow \infty$ and $\beta<1$, from \emph{Random Matrix Theory}, we have
\begin{equation}\label{radius1}
\rho(\gamma(\mathbf{H}^T\mathbf{H}-\mathbf{D}_{\mathbf{H}^T\mathbf{H}}))\to \beta+2\sqrt{\beta},
\end{equation}
for a finite $snr$. Then, from the second condition of \emph{Theorem 2}, we have the following corollary.

\textbf{\emph{Corollary 1:}}  \emph{For the Massive MU-MIMO system where $\beta= K/M$ is fixed, $K\to \infty$, and the transmitted symbols of the sources are i.i.d. with $\mathcal{N}^K(0,\sigma_x^2)$, the GMPID converges to the MMSE detection if $\beta<(\sqrt2-1)^2$.}

Let $\bigtriangleup \mathbf{e}(t)=\mathbf{e}^*-\mathbf{e}^v(t)$ be the mean deviation vector. From (\ref{p11}), we can get
\begin{equation}\label{conv_speed}
\Delta \mathbf{e}(t) = \gamma \left( {{\mathbf{H}^T}\mathbf{H} - {\mathbf{D}_{{\mathbf{{H}}^T}\mathbf{H}}}} \right)\Delta \mathbf{e}(t - 1).
\end{equation}
From (\ref{conv_speed}), we can see that the means converge to the fixing point with an exponential speed of the spectral radius $\rho \left( {\gamma ({\mathbf{H}^T}\mathbf{H} - {\mathbf{D}_{{\mathbf{H}^T}\mathbf{H}}})} \right)$, i.e., the smaller spectral radius is, the faster convergence speed it will have. Therefore, both the convergence condition and the convergence speed of GMPID can be improved by minimizing the spectral radius of the GMPID, which motivates us to modify the GMPID to get a better convergence condition and a better convergence speed.

\section{A New Fast-Convergence Detector SA-GMPID}
{As shown in the convergence analysis in Section III, the GMPID does not always converge to the optimal MMSE detection. The main reason is that the spectral radius of GMPID does not achieve the minimum value. Therefore, we propose a new \emph{scale-and-add} GMPID (SA-GMPID). The SA-GMPID is achieved by modifying the mean updates of GMPID with linear operators, that is \emph{i)} scaling the received $\mathbf{y}$ and the channel matrix $\mathbf{H}$ with a relaxation parameter $w$, i.e., ${\mathbf{H}'}=\sqrt{w}\mathbf{H}$ and $\textbf{\emph{y}}'=\sqrt{w}\textbf{\emph{y}}$, where $h'_{mk}=\sqrt{w}h_{mk}$ is an element of matrix $\mathbf{H}'$, and \emph{ii)} adding a new term $\!(\!w\!-\!1\!)e_{k \to m}^v\!(\!t\!-\!1\!)$ for the mean message update at each variable node. However, all the variance updates of SA-GMPID are kept the same, because we have proved in \emph{Theorem 1} that the variances converge to the exact MSE of the optimal MMSE detection. By doing so, we can optimize the relaxation parameter $w$ to minimize the spectral radius of SA-GMPID. As a result, the SA-GMPID will always converge to the optimal MMSE detection and have a faster convergence speed (see \emph{Theorem 3} and \emph{Corollary 2}). In addition, the SA-GMPID has the same complexity as the previous GMPID. In the following, we present the SA-GMPID.}\vspace{-0.35cm}

\subsection{Message Update at Sum Nodes of SA-GMPID}
The message update at the sum nodes (\ref{np1}) is changed to
\begin{equation}\label{FC2}
\left\{ \begin{array}{l}
e_{m \to k}^s(t) = {y'_m} - \sum\limits_{i \ne k} {{h_{mi}'}e_{i \to m}^v(t-1),} \\
v_{m \to k}^s(t)= \sum\limits_{i \ne k} {{h}_{mi}^2v_{i \to m}^v(t-1) + {{\sigma} _n^2}\;},
\end{array} \right.
\end{equation}
for $i,k \in \{ 1,2, \cdots ,K\}$ and $m \in \{ 1,2, \cdots ,M\}$.\vspace{-0.35cm}
\subsection{Message Update at Variable Nodes of SA-GMPID}
The message update of the variable nodes (\ref{var_mes}) is modified as
\begin{equation}\label{FC1}
\!\!\!\!\left\{\!\!\!\! {\begin{array}{*{20}{l}}
{v_{k \to m}^v(t) = {{( {\sum\limits_{i } {{h}_{ik}^2v_{i \to k}^{s{\;^{ - 1}}}(t) + \sigma _{{x_k}}^{ - 2}\;} } )}^{ - 1}},}\\
{e_{k \to m}^v\!(t) \!\!=\!\! v_{k \to m}^v(t)\!\!\sum\limits_{i} {{{h_{ik}'}}v_{i \to k}^{s{\;^{ - 1}}}\!(t)e_{i \to k}^s(t)\!-\!(\!w\!-\!1\!)e_{k \to m}^v\!(\!t\!-\!1\!),\;\;} }
\end{array}} \right.\quad
\end{equation}
for $k \in \{ 1,2, \cdots ,K\} $ and $ i,m \in \{ 1,2, \cdots ,M\}$.\vspace{-0.35cm}
\subsection{Decision and Output of SA-GMPID}
After several iterations between (\ref{FC1}) and ({\ref{FC2}}), output
\begin{equation}\label{FC3}
\!\left\{ \!\!\!\begin{array}{l}
{\sigma}^2_{\hat{x}_k} = {( {\sum\limits_{m } {{h}_{mk}^2v_{m \to k}^{s{\;^{ - 1}}}(t) + \sigma _{{x_k}}^{ - 2}\;} } )^{ - 1}},\\
\hat{x}_k \!=\! \!\sum\limits_{m } \!\left({\sigma}^2_{\hat {x}_k}{{h_{mk}'}v_{m \to k}^{s{\;^{ - 1}}}(t)e_{m \to k}^s(t)\!-\!\frac{w-1}{M}e_{k \to m}^v(t\!-\!1)}\right)\!,
\end{array} \right.
\end{equation}
where $k \!\in \!\{ 1,2, \cdots ,K\}$ and $ m \!\in\! \{ 1,2, \cdots ,M\}$. The detailed process of SA-GMPID is given in Algorithm 2.\vspace{-0.35cm}

\subsection{Variance Convergence of SA-GMPID}
As the variance updates of SA-GMPID are the same as those of the GMPID, the variances of SA-GMPID converge to the same values as those of the GMPID, i.e., $v_{k\to m}^v(t)$ and $v_{m\to k}^s(t)$ converge to $\hat{\sigma}^2$ and $\tilde{\sigma}^2$ respectively. When $\beta  < 1$, we have
 \begin{equation}\label{np7}
 \gamma={{{{\hat {\sigma} }^2}} \mathord{\left/
 {\vphantom {{{{\hat {\sigma} }^2}} {{{\tilde {\sigma} }^2}}}} \right.
 \kern-\nulldelimiterspace} {{{\tilde {\sigma} }^2}}}= \frac{1}{{{\rm{K}} + {{\sigma _n^2} \mathord{\left/
 {\vphantom {{\sigma _n^2} {{{\hat \sigma }^2}}}} \right.
 \kern-\nulldelimiterspace} {{{\hat \sigma }^2}}}}} \approx \left({M + sn{r^{ - 1}}} \right)^{ - 1}.
 \end{equation}

\begin{algorithm}[ht]
\caption{SA-GMPID Algorithm}
\begin{algorithmic}[1]
\State {\small{\textbf{Input:} {{$\mathbf{H}$, $\mathbf{V}_{\textbf{\emph{x}}}$, $\sigma^2_n$, $\epsilon>0$, $N_{ite}$}}, $\gamma$, calculate $w$, $\mathbf{H}^{(2)}$, $\mathbf{H}'$, $\textbf{\emph{y}}'$.
\vspace{0.1cm}
\State \textbf{Initialization:} {{$t\!=\!0$, $\mathbf{E}_{us}(0)\!=\!\left[0\right]_{K\times M}$}}, {{$\mathbf{V}_{us}(0)=\left[+\infty\right]_{K\times M}$.}}
\State \textbf{Do}
\State \vspace{-0.1cm}{\;\;\;$t\!=\!t\!+\!1$, $\widetilde{\mathbf{E}}_{us}(t)\!=\!\mathbf{E}_{us}(t).*\mathbf{H}'^T$\!, $\!\widetilde{\mathbf{V}}_{us}(t)\!=\!\mathbf{V}_{us}(t).\!*\!{\mathbf{H}^{(2)}}^{T}$,
\State  \vspace{-0.15cm}\[\;\begin{array}{l}\left[ \!\!\!\begin{array}{l}
{{\rm{\mathbf{E}}}_{su}}(t)\\
{\mathbf{V}_{su}}(t)
\end{array} \!\!\!\!\right] \!\!= \!\!\left[\!\!\! \begin{array}{l}
\quad\;\;\textbf{\emph{y}}' - dia{g^{ - 1}}\{ \mathbf{1}_{M\times K}\cdot{\widetilde{\mathbf{E}}_{us}}(t - 1)\} \\
\sigma_n^2\!\cdot\!{\mathbf{1}}_{M\times1}\!+\!dia{g^{ - 1}}\!\{ \mathbf{1}_{M\times K}\!\cdot\! {\widetilde{\mathbf{V}}_{us}}(t \!-\! 1)\}
\end{array} \!\!\!\!\right]\! \cdot\! {\mathbf{1}_{1 \times K}} \mathop {}\limits_{\mathop {}\limits_{\mathop {}\limits_{} } } \\
\qquad\qquad\quad-\left[ \begin{array}{l}
-\widetilde{\mathbf{E}}_{us}^T(t - 1)\\
\widetilde{\mathbf{V}}_{us}^T(t - 1)
\end{array} \right],\end{array}\qquad\qquad\qquad\]}
\vspace{-0.2cm}
\State \vspace{-0.3cm}\[\;\; \widetilde{\mathbf{W}}_{su}(t)\!=\!\mathbf{H}^{(2)}\!.*\mathbf{V}^{(-1)}_{su}\!(t), \widetilde{\mathbf{G}}_{su}\!(t)\!=\!\mathbf{H}'\!.*\mathbf{V}^{(-1)}_{su}\!(t).*\mathbf{E}_{su}(t),\qquad\]
\State \vspace{-0.35cm}\[ \begin{array}{l}
\left[\!\! \begin{array}{l}
{\mathbf{W}_{us}}(t)\\
{\mathbf{G}_{us}}(t)
\end{array} \!\!\!\right] \!\!=\!\! \left[ \!\!\!\begin{array}{l}
\mathbf{v}_{\textbf{\emph{x}}}^{(-1)}\!\! +\! dia{g^{ - 1}}\{ {{\mathbf{1}_{K \times M}} \!\cdot\! {{\widetilde{\mathbf{W}}_{su}}(t)}} \}\mathop {}\limits_{\mathop {}\limits_{} }\\
\quad\;\; dia{g^{ - 1}}\{ {{\mathbf{1}_{K \times M}} \cdot \widetilde{\mathbf{G}}_{su}(t)}\}
\end{array} \!\!\!\!\right] \!\cdot\! {\mathbf{1}_{1 \times M}},
\end{array}\qquad\qquad\qquad\]
\vspace{-0.3cm}
\State \vspace{-0.3cm}\[ \;\;\mathbf{V}_{us}\!(t)\!\!=\!\!\mathbf{W}_{us}^{(-1)}\!(t),\!  \mathbf{E}_{us}(t)\!\!=\!\!\mathbf{\mathbf{V}}_{us}\!(t).\!*\!\mathbf{G}_{us}\!(t)\!-\!(w\!-\!1)\mathbf{E}_{us}\!(t\!-\!1).\]
\State \textbf{While} \;{\small{$\left( \:|\mathbf{E}_{us}(t)-\mathbf{E}_{us}{(t-1)}|>\epsilon \;{\textbf{or}}\; t<N_{ite} \;\right)$}}
\State  \vspace{-0.3cm}{{\[  \begin{array}{l}{\mathbf{\sigma}^2_{\hat{\textbf{\emph{x}}}}} = {\left( {dia{g^{ - 1}}\{ {{\mathbf{1}_{K \times M}} \cdot {\widetilde{\mathbf{W}}_{su}}(t) } \}} \right)^{( - 1)}},\\
{\hat{\textbf{\emph{x}}}} = {{\mathbf{\sigma}^2_{\hat{\textbf{\emph{x}}}}}}.*dia{g^{ - 1}}\{ {\mathbf{1}_{K \times M}} \cdot (\widetilde{\mathbf{G}}_{su}(t)-\frac{w-1}{M}\mathbf{E}_{us}^T(t-1)) \}. \end{array}\]}}
\State \textbf{Output:}  ${\hat{\textbf{\emph{x}}}}$ and ${\mathbf{\sigma}^2_{\hat{\textbf{\emph{x}}}}}$ }}.
\end{algorithmic}
\end{algorithm}\vspace{-0.35cm}
\subsection{Mean Convergence of SA-GMPID}
Now, we discuss the mean convergence of SA-GMPID. From (\ref{FC1}) and (\ref{FC2}), we have
\begin{equation}\label{FC6}
\begin{array}{l}
\!\!\!\!\!e_{k \to m}^v(t)\!=\! v_{k \to m}^v(t)\!\sum\limits_i \!{{{h_{ik}'}}v_{i \to k}^{s{^{ - 1}}}(t)( {{{y_i}'} \!\!-\!\!\! \sum\limits_{j \ne k} \!{{{h_{ij}'}}e_{j \to i}^v(t\! -\! 1)} } )}  \\
\qquad\qquad\;- (w - 1)e_{k \to m}^v(t - 1).
\end{array}
\end{equation}
From \emph{Proposition 3}, when $t$ is large enough, (\ref{FC6}) can be rewritten to
\begin{equation}\label{FC7}
e_{k \to m}^v(t) \!=\! \gamma\!\!\sum\limits_{i} \!{{h'_{ik}}( {{y'_i} \!\!-\!\!\!  \sum\limits_{j \ne k}\! {{h'_{ij}}e_{j \to i}^v(t \!-\! 1)} }  \!)\!-\!(w\!-\!1)e_{k \to m}^v\!(t\!-\!1)}.
\end{equation}
Then, we can get $e_{k \to m}^v(t)=e^v_{k}(t)$, for $k \in \{ 1, \cdots ,K\}$ and $m \in \{ 1, \cdots ,M\}$. Thus, the matrix form of (\ref{FC7}) can be rewritten to
\begin{equation}\label{FC8}
\mathbf{e}^v(t) \!=\!  \gamma{{\mathbf{H}'^T}\textbf{\emph{y}}' \!-\! \left[\gamma\left({\mathbf{H}'^T}\mathbf{H}'\!\!-\!\!\mathbf{D}_{{\mathbf{H}'^T}\mathbf{H}'}\right)\!+\!(w\!-\!1)\mathbf{I}_{K}\!\right]\! \mathbf{e}^v\!(t\!-\!1)}.
\end{equation}
Based on this analysis, we can have the following theorem.

\textbf{\emph{Theorem 3:}}  \emph{For the Massive MU-MIMO system where $\beta= K/M$ is fixed, $K\to \infty$, and the transmitted symbols of the sources are i.i.d. with $\mathcal{N}^K(0,\sigma_x^2)$, the SA-GMPID converges to the MMSE detection if the relaxation parameter $w$ satisfies $0<w<2/\lambda_{max}^\mathbf{A}$, where $\lambda_{max}^\mathbf{A}$ is the largest eigenvalue of matrix $\mathbf{A}=\gamma\left(\mathbf{H}^T\mathbf{H}-\mathbf{D}_{\mathbf{H}^T\mathbf{H}}\right)+\mathbf{I}_{K}$.}

\begin{IEEEproof}
From (\ref{FC8}), we can see that SA-GMPID is an equivalent classical iterative algorithm in (\ref{a1}). Thus, according to \emph{Proposition 1}, the SA-GMPID converges if
\begin{equation}\label{FC9}
\rho\left(\gamma\left({\mathbf{H}'^T}\mathbf{H}'-\mathbf{D}_{{\mathbf{H}'^T}\mathbf{H}'}\right)+(w-1)\mathbf{I}_{K}\right)<1, \end{equation}
i.e., $\rho\left(\mathbf{I}-w\mathbf{A}\right)<1$, and $\mathbf{A}=\gamma\left(\mathbf{H}^T\mathbf{H}-\mathbf{D}_{\mathbf{H}^T\mathbf{H}}\right)+\mathbf{I}_{K}$.
When $K, M \to \infty$, the smallest and largest eigenvalues ${\lambda}_{max}^\mathbf{A}$ and ${\lambda}_{min}^\mathbf{A}$ of matrix $\mathbf{A}$ are given by \cite{verdu2004}
\begin{equation}\label{eigv1}
{\lambda}_{min}^\mathbf{A}\!=\!1+\gamma M\!{(\!( {\!1 \!- \!\sqrt {\beta} } )^2\!\!\!-\!1)} ,{\lambda}_{max}^\mathbf{A}\!=\!1+\gamma M{\!(\!( {\!1\! \!+ \sqrt {\beta} } )^2\!\!\!-\!1\!)}.
\end{equation}
When $\beta<1$, from (\ref{p7}), we have $M\gamma\to 1$. Then we can get ${\lambda}_{max}^\mathbf{A}>{\lambda}_{min}^\mathbf{A}>0$ from (\ref{eigv1}), which means that $\mathbf{A}$ is strictly positive definite. Therefore, the condition (\ref{FC9}) can always be satisfied if the relaxation parameter $w$ satisfies $0<w<2/\lambda_{max}^\mathbf{A}$. In addition, we can always find such a $w$ that satisfies $0<w<2/\lambda_{max}^\mathbf{A}$ because ${\lambda}_{max}^\mathbf{A}>0$.

Next, we will show that the SA-GMPID converges to the MMSE detection. From \emph{Proposition} 1, we know that $\mathbf{e}^v(t)$ converges to
\begin{equation}\label{FC13}
{\mathbf{e}^*} = {\left( {snr^{-1}{\mathbf{I}_K} + {\mathbf{H}^T}\mathbf{H}} \right)^{ - 1}}{\mathbf{H}^H}\textbf{\emph{y}},
\end{equation}
which is the same as (\ref{p12}). Following the proof of \emph{Theorem 1}, it can be shown that the SA-GMPID converges to the MMSE detection.
\end{IEEEproof}

Now, we discuss the value of $w$. The relaxation parameter $w$ can be optimized by \cite{Gao2014}
\begin{equation}\label{FC10}
w=2/({\lambda}_{min}^\mathbf{A}+{{\lambda}_{max}^\mathbf{A}}).
\end{equation}
It minimizes the spectral radius of $\mathbf{I}_{K}-w\mathbf{A}$ and we get $\rho_{min}(\mathbf{I}_{K}-w\mathbf{A})= \frac{{\lambda}_{max}^\mathbf{A}-{\lambda}_{min}^\mathbf{A}}{{\lambda}_{max}^\mathbf{A}+{\lambda}_{min}^\mathbf{A}}<1$ . From (\ref{eigv1}), we have $w = {1 \mathord{\left/
 {\vphantom {1 {\left( {1 + \gamma M \beta} \right)}}} \right.
 \kern-\nulldelimiterspace} {\left( {1 + \gamma M \beta} \right)}}$, where $\beta=K/M$. When $\beta<1$, from (\ref{np7}), we get $M\gamma\to 1$. Thus, we can simplify (\ref{FC10}) to
 \begin{equation}\label{FC12}
 w={1 \mathord{\left/
 {\vphantom {1 {(1 + \beta )}}} \right.
 \kern-\nulldelimiterspace} {(1 + \beta )}},\quad \rho_{min}(\mathbf{I}_{K}-w\mathbf{A})\approx\frac{2\sqrt{\beta}}{1+\beta}<1.
 \end{equation}
 Therefore, the SA-GMPID converges to MMSE with a speed $\left(\frac{2\sqrt{\beta}}{1+\beta}\right)^t$, where $t$ denotes the number of iterations. Comparing with (\ref{radius1}), we have the following corollary.

\textbf{\emph{Corollary 2:}} \emph{For the Massive MU-MIMO system where $\beta\!\!=\!\! K/M$ is fixed, $K\!\!\to \!\!\infty$, and the transmitted symbols of the sources are i.i.d. with $\mathcal{N}^K(0,\sigma_x^2)$, the SA-GMPID has a faster convergence speed than the GMPID.}

\textbf{\textit{Remark 4}}: When $K$ and $M$ are finite, the smallest and largest eigenvalues of the matrix $\mathbf{A}$ given by (\ref{eigv1}) are inaccurate, which may impact the convergence and MSE performance of the SA-GMPID. To improve the convergence speed and the robustness of the algorithm, we can set the relaxation parameter as $w=2/\lambda^*_\mathbf{A}$, where $\lambda^*_\mathbf{A}$ is an upper bound of the eigenvalues of matrix $\mathbf{A}$ \cite{Varga2009}, which introduces a little bit more calculations.
\begin{equation}\label{AFC10}
{\lambda _{\max }} \le \lambda^*_\mathbf{A}= \min \{ \mathop {max}\limits_{1 \le i \le K} \sum\limits_{j = 1}^K {\left| {{a_{ij}}} \right|} ,\mathop {max}\limits_{1 \le j \le K} \sum\limits_{i = 1}^K {\left| {{a_{ij}}} \right|} \}.
\end{equation}

{\textbf{\textit{Remark 5}}: The proof of \emph{Theorem 3} and the spectral radius analysis show the detailed process of the SA-GMPID design. The main reason that we propose the SA-GMPID is that its spectral radius is minimized after the modification of GMPID. As a result, the convergence prerequisite and convergence speed of the SA-GMPID are significantly improved.}
{\newcommand{\tabincell}[2]{\begin{tabular}{@{}#1@{}}#2\end{tabular}}
\begin{table*}[t]
\renewcommand{\arraystretch}{1.8}
\caption{Complexity comparison of the different detection algorithms for MU-MIMO system}
\label{table2}
\centering
\begin{tabular}{|c|c|c|}
\hline
 \tabincell{c}{NONITERATIVE\vspace{-0.35cm}\\ DETECTION ALGORIHMS} & IF \& GMP & MMSE \\
\hline
  COMPLEXITY & $\mathcal{O}(M^2K+M^3)$ &  $\mathcal{O}(\min\{MK^2+K^3,\;KM^2+M^3\})$\\
\hline
\tabincell{c}{ITERATIVE\vspace{-0.35cm}\\DETECTION ALGORIHMS}  & \tabincell{c}{Jacobi \& GaBP \vspace{-0.35cm}\\ \& Richarson}& \tabincell{c}{ GMPID \& SA-GMPID }\\
\hline
  COMPLEXITY& $\mathcal{O}(MK^2+K^2N_{ite})$\!\!&$\mathcal{O}({MK}N_{ite})$  \\
\hline
\end{tabular}
\end{table*}}

{ \subsection{Relationship with the IF and GMP Algorithms}
\subsubsection{Inverse Filter}
Inverse Filter (IF) \cite{tse2005} is also known as interference nulling, decorrelator or zero-forcing receiver. It is given by
\begin{equation}\label{IF1}
\begin{array}{l}
\tilde {\textbf{\emph{x}}} \!=\! {({\mathbf{H}^T}\mathbf{H})^{ - 1}}{\mathbf{H}^T}\textbf{\emph{y}} = \textbf{\emph{x}} + {({\mathbf{H}^T}\mathbf{H})^{ - 1}}{\mathbf{H}^T}\textbf{\emph{n}}=\textbf{\emph{x}} + {\textbf{\emph{n}}}',\\
\end{array}
\end{equation}
where $ {\textbf{\emph{n}}'}\sim \mathcal{N}(\textbf{0}, \mathbf{V}_{ {\textbf{\emph{n}}'}})$ with ${\mathbf{V}_{ {{\textbf{\emph{n}}}'}}} = (\mathbf{H}^T\mathbf{H})^{-1}\sigma^2_n$. By combining the estimated distribution $\mathcal{N}(\tilde {\textbf{\emph{x}}},\mathbf{V}_{ {{\textbf{\emph{n}}}'}})$ with  $\mathcal{N}(\textbf{0},\mathbf{V}_{\textbf{\emph{x}}})$ according to the first message update rule in Fig. \ref{update_rule}, we get
\begin{equation}\label{mmse8}
{{\hat {\textbf{\emph{x}}}}} = \sigma _{{{n}}}^{ - 2}\mathbf{V} _{{{\hat {\textbf{\emph{x}}}}}}\mathbf{H}^T\textbf{\emph{y}},
\end{equation}
where $\mathbf{V} _{{{\hat {\textbf{\emph{x}}}}}} = (\sigma _{{{n}}}^{ - 2}\mathbf{H}^T\mathbf{H}+\mathbf{V} _{{{ {\textbf{\emph{x}}}}}}^{-1})^{-1}$, which is the same as the MMSE detection.

\subsubsection{Gaussian Message Passing Algorithm}
Gaussian Message Passing (GMP) algorithm \cite{Forney2001,kschischang2001,Loeliger2004,Loeliger2002,Loeliger2006, Guo2008} considers the fourth message update rule for the matrix multiplication constraint in Fig. \ref{update_rule}. Replacing matrix $\mathbf{A}$ with the channel matrix $\mathbf{H}$, we have
\begin{equation}\label{GMP1}
\left\{ \begin{array}{l}
\mathbf{W}_{\tilde {\textbf{\emph{x}}}} = {\mathbf{H}^T}\mathbf{W}_\textbf{\emph{y}}^{in}\mathbf{H},\\
\mathbf{W}_{\tilde {\textbf{\emph{x}}}}\tilde{\textbf{\textbf{\emph{x}}}} = {\mathbf{H}^T}\mathbf{W}_\textbf{\emph{y}}^{in}\textbf{\emph{m}}_\textbf{\emph{y}}^{in},
\end{array} \right.
\end{equation}
where $\mathbf{W}_\textbf{\emph{y}}^{in}\!\!=\!\sigma_n^{-2}\mathbf{I}_M$ and $\textbf{\emph{m}}_\textbf{\emph{y}}^{in}\!\!=\!\textbf{\emph{y}}$.
Thus, we get a Gaussian estimation $\tilde {{\textbf{\emph{x}}}}$ and the estimation deviation matrix $\mathbf{W}_{\tilde {\textbf{\emph{x}}}}^{-1}$. By combining the estimated $\mathcal{N}(\tilde {\textbf{\emph{x}}},\mathbf{W}_{\tilde {\textbf{\emph{x}}}}^{-1})$ with  $\mathcal{N}(\textbf{0},\mathbf{V}_{\textbf{\emph{x}}})$, we get
\begin{equation}\label{GMP2}
{{\hat {\textbf{\emph{x}}}}} = \sigma _{{{n}}}^{ - 2} \mathbf{V} _{{{\hat {\textbf{\emph{x}}}}}}\mathbf{H}^T\textbf{\emph{y}}= \sigma _{{{n}}}^{ - 2}(\sigma _{{{n}}}^{- 2}\mathbf{H}^T\mathbf{H}+\mathbf{V} _{{{ {\textbf{\emph{x}}}}}}^{-1})^{-1}\mathbf{H}^T\textbf{\emph{y}},
\end{equation}
where $\mathbf{V} _{{{\hat {\textbf{\emph{x}}}}}} \!\!=\!\! (\sigma _{{{n}}}^{- 2}\mathbf{H}^T\mathbf{H}\!+\!\mathbf{V} _{{{ {\textbf{\emph{x}}}}}}^{-1})^{-1}$, which is the same as the MMSE detection. Hence, we have the following proposition.

\textbf{\textit{Proposition 4:}} \emph{If the distributions of the sources are known and are combined with the outputs of IF and GMP detector, the IF and GMP algorithms are equivalent to the MMSE algorithm for the uplink MU-MIMO detection.}

It means that the proposed GMPID also converges to the modified IF and GMP algorithms under the message update rule. It should be noted that, in general, the IF, MMSE detector and GMP algorithm are not equivalent. Their equivalence here is based on the condition that all the detectors are combined with source distributions according to the message update rules. For simplicity, we call all these three detections MMSE detection in this paper.}

{\subsection{Complexity Comparison of Different Detectors}
Table I compares the computational complexity between the GMPID and the other detection algorithms. The complexities of GMP detector and IF are both $\mathcal{O}(M^3+KM^2)$, where $\mathcal{O}(M^3)$ arises from the matrix inversion and $\mathcal{O}(KM^2)$ arises from calculation of $\mathbf{H}^H\mathbf{H}$, i.e., matrix multiplication. The complexity of the MMSE detector is $\mathcal{O}(\{K^3+MK^2,\;M^3+KM^2\})$, which is given in Section II-E. Similarly, the complexity in each iteration of the above classical iterative algorithms is $O(K^2)$, and the matrix calculation $\mathbf{A} = \mathbf{I}_K + snr\mathbf{H}^T\mathbf{H}$ costs $O(MK^2)$ operations before the iteration. So the total complexity of the classical iterative algorithm is $O(MK^2+K^2N_{ite})$.

It should be pointed out that the computational complexities of GMPID and SA-GMPID algorithms are almost the same, i.e., $\mathcal{O}(MKN_{ite})$ (see Section III. E). Besides, if the channel is sparse, the complexity of SA-GMPID (or GMPID) can be further reduced to $\mathcal{O}(N_\mathbf{H}N_{ite})$, where $N_\mathbf{H}$ is the number of nonzero elements in channel matrix $\mathbf{H}$. However, for the IF, GMP, MMSE, Jacobi, GaBP and Richardson algorithms, their complexities will not change with the sparsity of the channel matrix, as the sparsity is destroyed after calculating $\mathbf{H}^T\mathbf{H}$.}

\section{Simulation Results}
\begin{figure*}[t]
  \centering
  \includegraphics[width=15cm]{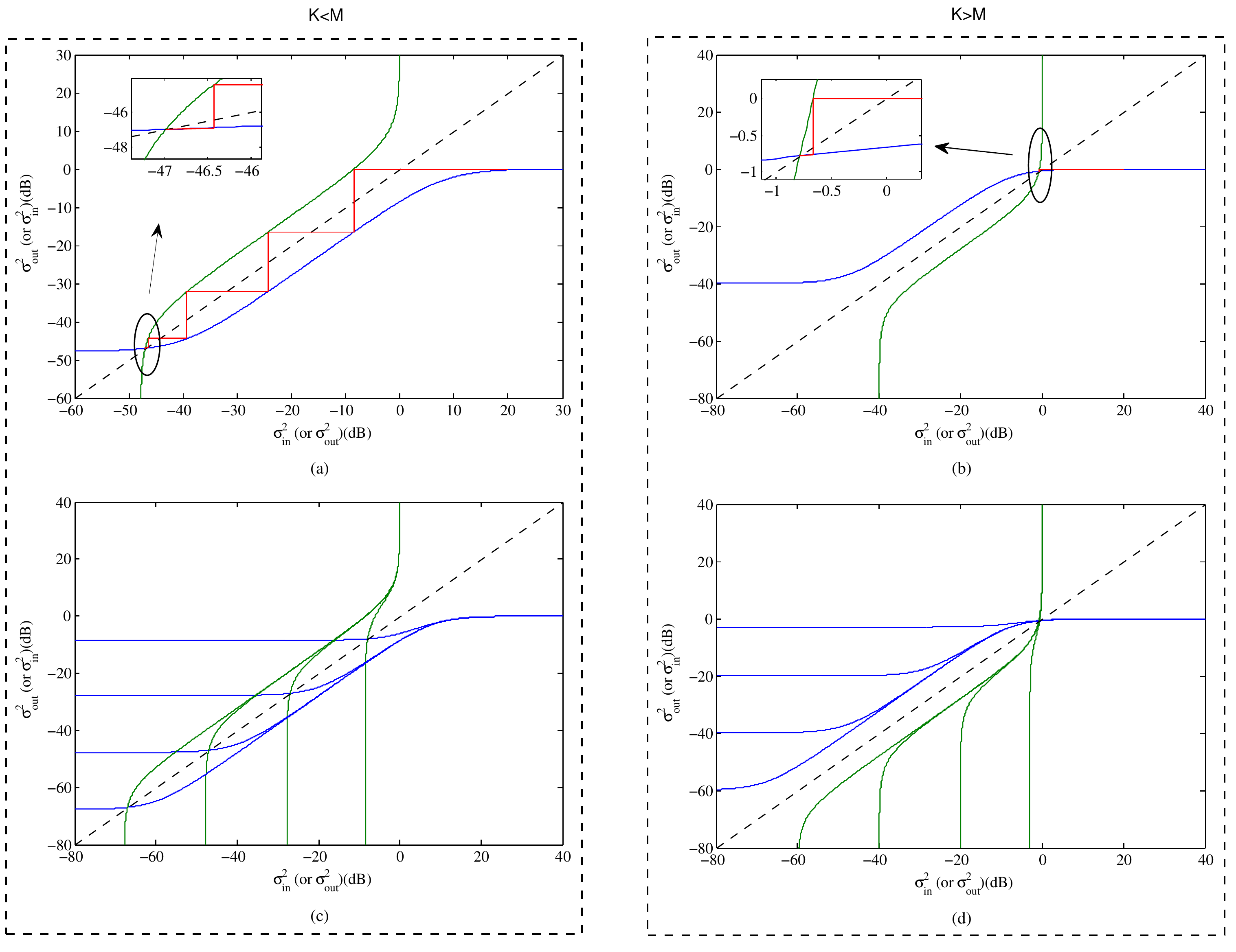}\\
  \caption{MSET chart analysis of the GMPID. In sub-figure (a), $K=100$, $M=600$, $\beta=1/6$ and $SNR=20$(dB). In sub-figure (b), $K=600$, $M=100$, $\beta=6$ and $SNR=20$(dB). In sub-figure (c), $K=100$, $M=600$, $\beta=1/6$ and $SNR=-20:20:40$(dB). In sub-figure (d), $K=600$, $M=100$, $\beta=6$ and $SNR=-20:20:40$(dB).
}\label{f5}\vspace{-0.3cm}
\end{figure*}

In this section, we present the numerical simulation results of the proposed detectors for MU-MIMO system with Gaussian sources. We assume that the sources are i.i.d. with $\mathcal{N}^K(0,1)$ and the entries of the channel matrix $\mathbf{H}$ are i.i.d. with normal distribution $\mathcal{N}^{M\times K}(0,1)$. In the following simulations, $N_{ite}$ denotes the number of iterations, $SNR=\dfrac{1}{\sigma^2_n}$ is the signal-to-noise ratio and $MSE=\frac{1}{K}\cdot E\left[\|\textbf{\emph{x}}-\hat{\textbf{\emph{x}}}\|^2\right]$ denotes the averaged mean squared error between the estimation and the transmitted sources. All the simulations are repeated with 500 random realizations.

\begin{figure}[t]
  \centering
  \includegraphics[width=8.5cm]{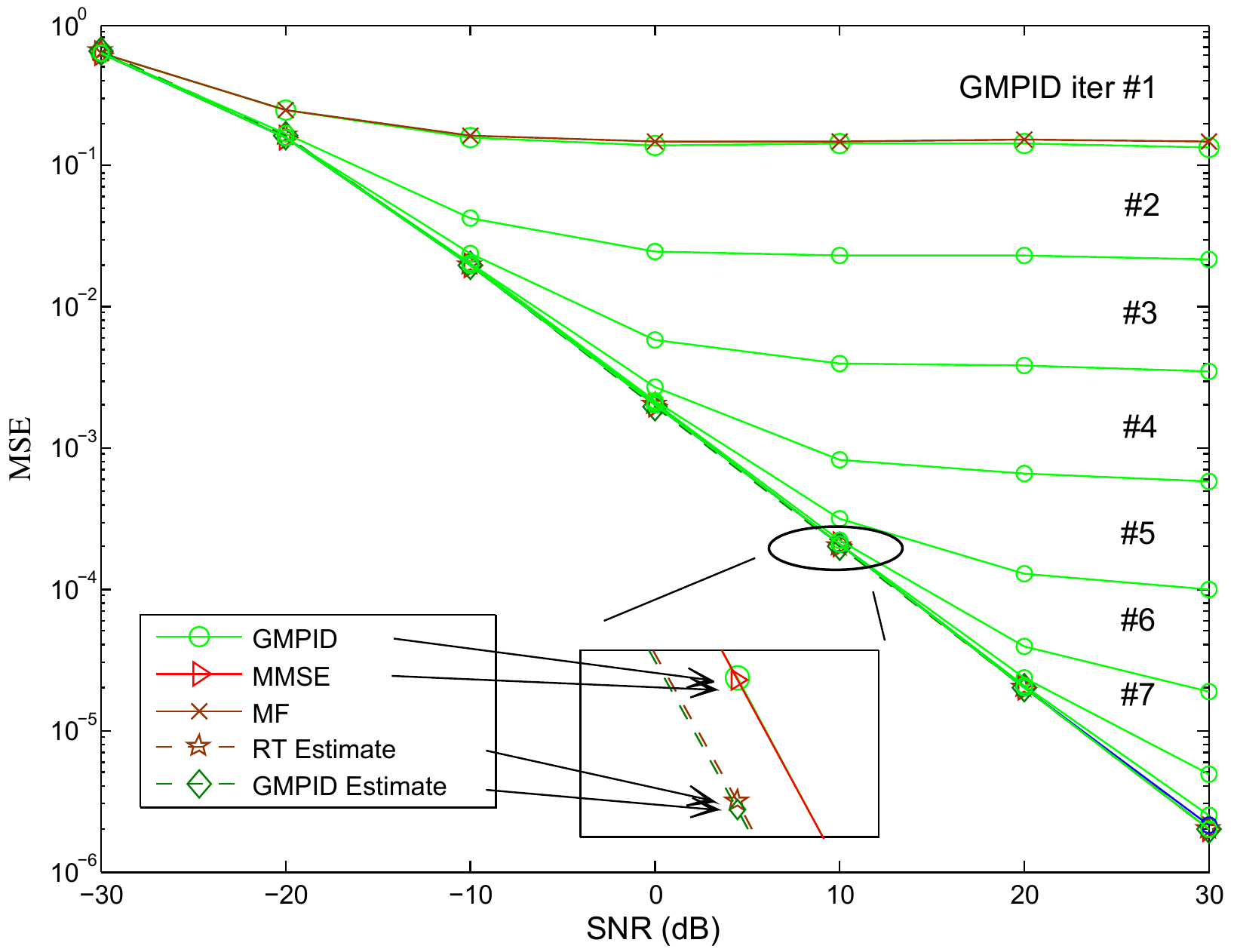}\\\vspace{-0.3cm}
  \caption{Performance comparison between the matched filter, MMSE detection and GMPID with 10 iterations; the performance estimates of the MMSE detection and the proposed GMPID for the MU-MIMO system with $K=100$, $M=600$ and $\beta=1/6$. }\label{f6}\vspace{-0.5cm}
\end{figure}
\begin{figure}[ht]
  \centering
  \includegraphics[width=9.2cm]{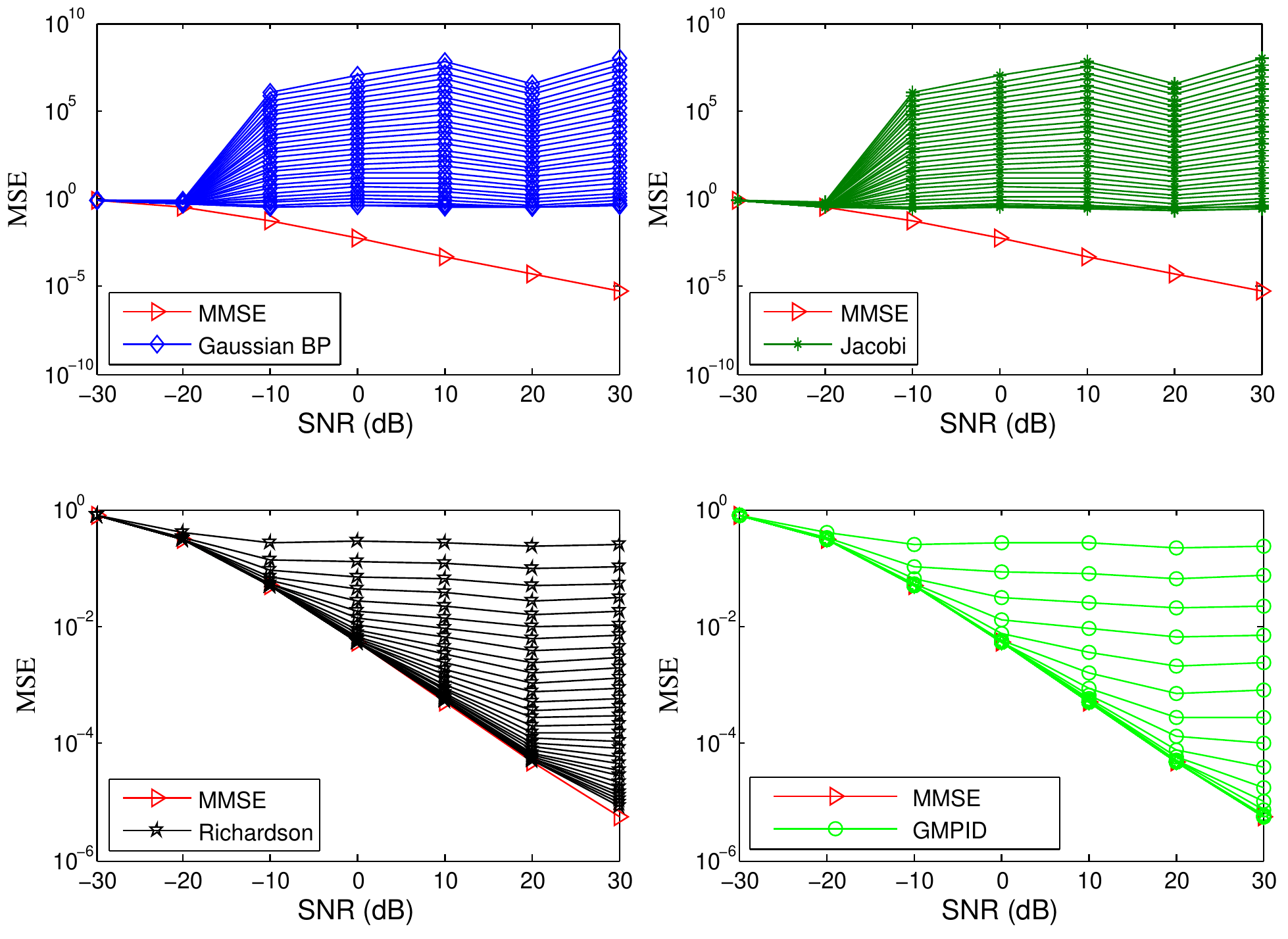}\\\vspace{-0.3cm}
  \caption{Performance comparison between GMPID and the other iterative algorithms: Gaussian BP, Jacobi and Richardson algorithm. The simulations are for $100\times300$ MU-MIMO system with $\beta=1/3$ and $1\sim30$ iterations.}\label{f7}\vspace{-0.3cm}
\end{figure}

{In Fig. \ref{f5}, we track the input and output MSE (or the input and output message variances: $\sigma^2_{in}$ and $\sigma^2_{out}$) during the iteration process of the GMPID algorithm. In this paper, the performance of the detectors is measured by the MSE instead of the mutual information. Therefore, the ``MSE Transfer (MSET) chart" is used to analyse the system performance, which is analogous to the EXIT chart \cite{Bhattad2007}. In fact, the MSET chart can be transformed into the EXIT chart based on the relationship between the mutual information and MSE \cite{Bhattad2007, Guo2005}.} Fig. \ref{f5}(a) and Fig. \ref{f5}(c) are considered for the underloaded $100\times600$ MU-MIMO system with $\beta=1/6$, and Fig. \ref{f5}(b) and Fig. \ref{f5}(d) are considered for the overloaded $600\times100$ MU-MIMO system with $\beta=6$. The SNR of the MSET chart curves in Fig. \ref{f5}(a) and Fig. \ref{f5}(b) is $20$ dB, and in Fig. \ref{f5}(c) and Fig. \ref{f5}(d), the SNRs of the MSET chart curves are $-20:20:40$ dB. Fig. \ref{f5}(a) and Fig. \ref{f5}(c) show the process that the variance decreases with increasing number of iterations. We can see that when $K<M$, the GMPID often converges to a fix point with a low variance (MSE). The variance of the converged point will be smaller with a higher SNR but it needs more iterations to reach there (slower convergence speed). From Fig. \ref{f5}(b) and Fig. \ref{f5}(d), we can see that when $K>M$, the GMPID converges to a fix point with a high variance (MSE) which is close to the prior variance of the sources, and the fix point moves slowly with increasing SNR. This is the reason why we only consider the case $\beta<1$ in this paper. However, in the case $\beta>1$, it has a very fast convergence speed (no more than 10 iterations in general). Besides, our simulation results also show that the larger the difference of $K$ and $M$, the larger the gap of the transfer curves, i.e., the GIMP converges faster.
\begin{figure*}[t]
  \centering
  \includegraphics[width=16.0cm]{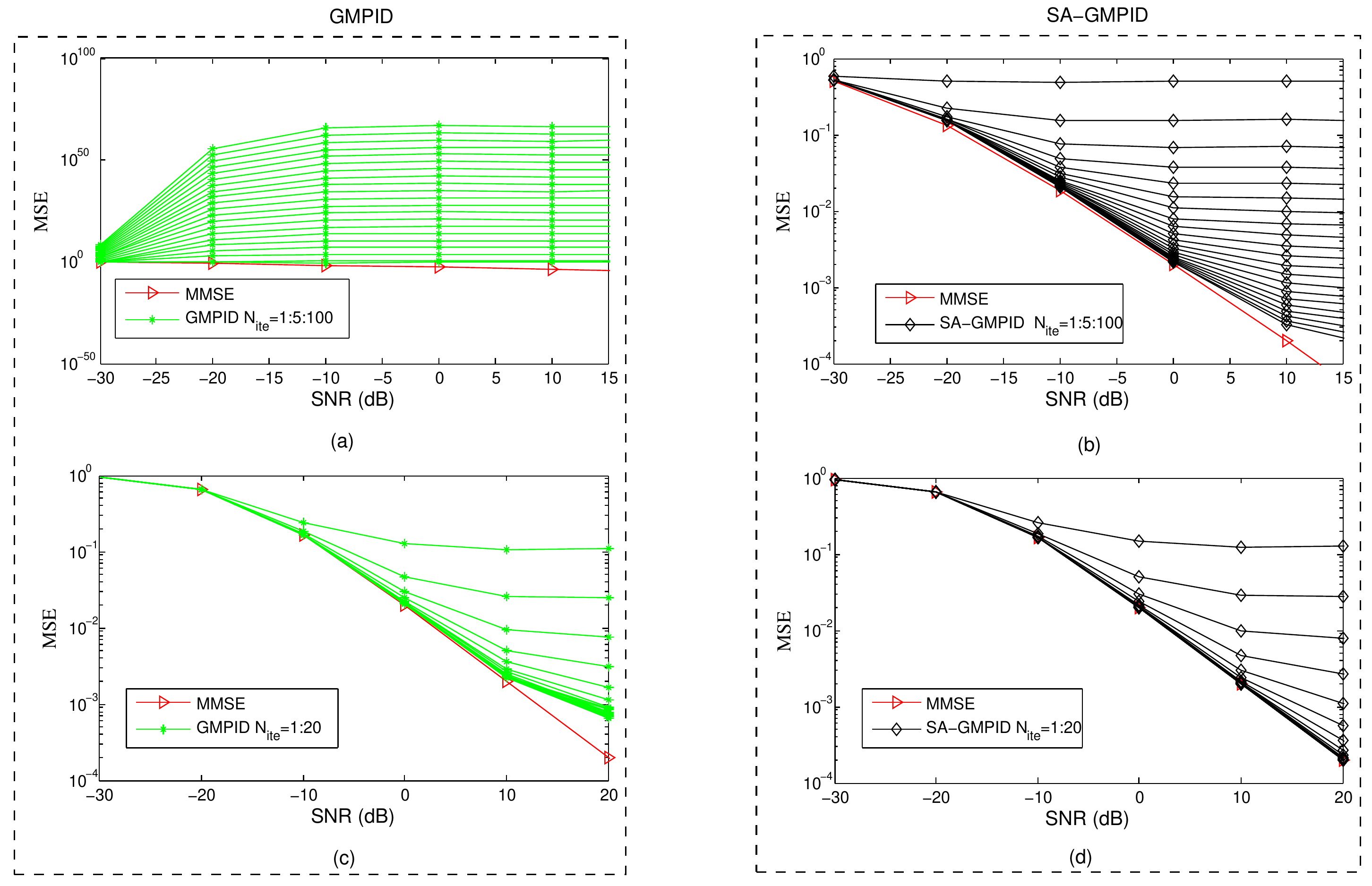}\\\vspace{-0.1cm}
  \caption{Performance comparison between GMPID and SA-GMPID (proposed) for the MU-MIMO system. In sub-figures (a) and (b), $K=1000$, $M=1500$, $\beta=2/3$. In sub-figures (c) and (d), $K=10$, $M=60$, $\beta=1/6$. }\label{f8}\vspace{-0.3cm}
\end{figure*}
Fig. \ref{f6} presents the averaged MSE performances and the performance estimates (see {\ref{p6}}) of the different detection algorithms for the MU-MIMO system with $K=100$, $M=600$ and $\beta=1/6$. The curve \emph{RT Estimation} is the MMSE detection performance estimate given by the \emph{Random Matrix Theory}, and the curve \emph{GMPID Estimation} is the GMPID performance estimate given by (\ref{p6}). It can be seen that the performance of MF (Matched Filter) is poor, while the other curves are almost completely overlapped. These verify that 1) the proposed GMPID converges fast (only 10 iterations) to the MMSE detection (\emph{Theorem 1}), 2) the performance estimates made by \emph{Random Matrix Theory} and GMPID match well to the simulation results (\emph{Proposition 1}). These results agree with the MSET chart analysis in Fig. \ref{f5}.

Fig. \ref{f7} gives the averaged MSE performance and convergence comparisons between the GMPID and the other iterative MU-MIMO detection algorithms, namely Jacobi, Gaussian BP and Richardson, where $K=100$, $M=300$ and $\beta=1/3$. We can see that the GMPID converges to the MMSE detection faster than the other three algorithms. Furthermore, the GMPID is convergent even when the Jacobi algorithm and Gaussian BP algorithm are divergent.

Fig. \ref{f8} gives the averaged MSE performance comparison between the GMPID and SA-GMPID for the cases that $\beta=2/3$, $K=1000$, $M=1500$ with $100$ iterations (Fig. \ref{f8}(a) and Fig. \ref{f8}(b)) and $\beta=1/6$, $K=10$, $M=60$ with 20 iterations (Fig. \ref{f8}(c) and Fig. \ref{f8}(d)) respectively. From Fig. \ref{f8}(a), we can see that GMPID diverges when $\beta\to1$. In contrast, Fig. \ref{f8}(b) shows that the SA-GMPID converges to the MMSE detection with increasing number of iterations. This verifies our analysis result in \emph{Theorem 2}. Furthermore, Fig. \ref{f8}(c) and Fig. \ref{f8}(d) show that 1) SA-GMPID converges to the MMSE detection faster than GMPID (\emph{Corollary 2}), 2) the proposed theoretical results even work for MU-MIMO systems with a small number of antennas and users ($K=10$, $M=60$).

Table \ref{table} concludes the convergence comparison of the different MU-MIMO detection algorithms, where ``C" (or ``D") denotes converging to (or diverging from) the MMSE detection and $``+"$ (or $``-"$) denotes right limit (left limit). It shows that 1) all the iterative algorithms are convergent when $\beta<(\sqrt{2}-1)^2$, 2) the Jacobi and GaBP algorithms are divergent when $\beta>(\sqrt{2}-1)^2$, 3) GMPID and SA-GMPID are still convergent when $\beta$ is close to $(\sqrt{2}-1)^2_+$ and 4) Richardson algorithm and SA-GMPID are convergent even when $\beta$ is close to $1$.

Fig. \ref{f9} illustrates the computational complexity comparison between the different detection algorithms with respect to the MSE (determined by the $SNR$ of the system) for $500\times3500$ MU-MIMO system with $\beta=1/7$. The relative error of MSE of each iterative algorithm is less than 0.1. We can see that the complexity of the SA-GMPID detector increases with decreasing MSE, i.e., the lower MSE we need, the higher complexity it costs. It also shows that 1) the MMSE detector has a constant and the highest complexity, 2) the classical iterative algorithms like GaBP, Jacobi and Richardson algorithms always have much higher complexities than the GMPID, 3) the SA-GMPID has a significant reduction of system complexity compared with the GMPID. Therefore, the proposed SA-GMPID has a good convergence performance and much lower computational complexity. In addition, the SA-GMPID can achieve more improvement in complexity and performance with larger number of users or antennas.

\newcommand{\tabincell}[2]{\begin{tabular}{@{}#1@{}}#2\end{tabular}}
\begin{table}[t]
\renewcommand{\arraystretch}{1.5}
\caption{Convergence comparison between SA-GMPID, GMPID, Jacobi, GaBP and Richardson algorithms. The ``C" and ``D" denote converging to and diverging from the MMSE detection respectively, and $``+"$ and $``-"$ denote the right and left limit respectively.}
\label{table}
\centering
\begin{tabular}{|c|c|c|c|c|c|}
\hline
 Figure & $\beta$ & \tabincell{c}{ Jacobi\\ \& GaBP} &  \tabincell{c}{GMPID }  & \tabincell{c}{Richardson \&\\ SA-GMPID} \\
\hline
  Fig. \ref{f6} &$\beta<(\sqrt{2}-1)^2$ & C & C & C \\
\hline
 Fig. \ref{f7} & $\beta\to(\sqrt{2}-1)^2_+$ & D & C & C \\
\hline
 Fig. \ref{f8} & $\beta\to1_-$ & D & D & C \\
\hline
\end{tabular}
\end{table}

{ To show that the proposed SA-GMPID also works for the discrete modulated systems, Fig. \ref{f10} presents the BER performances of the SA-GMPID and MMSE detection for the practical $10\times60$ and $10\times30$ MU-MIMO communication systems transmitting digital modulation waveforms. For the $10\times30$ MU-MIMO communication system with $\beta=1/3$, the original GMPID is divergent. In these two systems, each user is encoded with a Turbo Hadamard channel code \cite{Ping2003_2,Ping2004}(3 component codes, Hadamard order: 5, spread length: 4), where the code rate is 0.01452 bits/symbol and the code length is $2.82\times 10^5$. A 10-bit superposition coded modulation \cite{Wachsmann1999, Gadkari1999} is employed for each user to produce Gaussian like transmitting signals. Hence, the transmitting length of each user is $2.82\times 10^4$, the rate of each user is $R_u=0.1452$ bits/symbol, and the system sum rate is 1.452 bits per channel use. $E_b/N_0$ is calculated by $E_b/N_0\!=\!\dfrac{P_u}{2R_u\sigma^2_n}$, where $P_u\!=\!1$ is the power of each user, and $\sigma^2_n$ is the variance of the Gaussian noise. The Shannon limit of $10\!\!\times\!\!60$ MU-MIMO system is $E_b/N_0\!\!=\!\!-16.37$dB, and $10\!\times\!30$ MU-MIMO system is $E_b/N_0\!\!=\!\!-13.88$dB. The base station recovers the messages of all users by iterative multi-user decoding between the detector and separate user decoders. The detector and user decoders exchange extrinsic soft information (means and variances) of the transmitting signals with each other. It means that the MMSE detector has to perform one detection in each iteration, which increases the complexity of MMSE detector to $\mathcal{O}(\min\{(MK^2\!\!+\!\!K^3)N_{ite},\;(KM^2\!\!+\!\!M^3)N_{ite}\})$ that is $N_{ite}$ times the original detection complexity. Fig. \ref{f10}(a) shows that for the $10\times60$ MU-MIMO system with the channel coding at each user, it only needs 2 iterations for the proposed SA-GMIPD to converge to the MMSE detection, and after 5 iterations the BER performance is 2.37dB away from the Shannon limit. Fig. \ref{f10}(b) shows that for the $10\!\!\times\!\!30$ MU-MIMO system with the channel coding at each user, it only needs 5 iterations that the proposed SA-GMIPD converges to the MMSE detection, and 5 iterations is good enough to achieve a BER performance that is 2.9dB away from the Shannon limit. This phenomenon is consistent with our theoretic analysis which points out that the proposed SA-GMIPD can always converge to the MMSE detection quickly, even when $\beta\!=\!1/3\!>\!(\sqrt{2}\!-\!1)^2\!$. It should be noted that we do not consider the matching design between the detector and the channel codes in this system. If the detector and channel codes were globally optimized, the BER performance could be further improved, but this is beyond the scope of this paper.}
\begin{figure}[t]
  \centering
  \includegraphics[width=9.0cm]{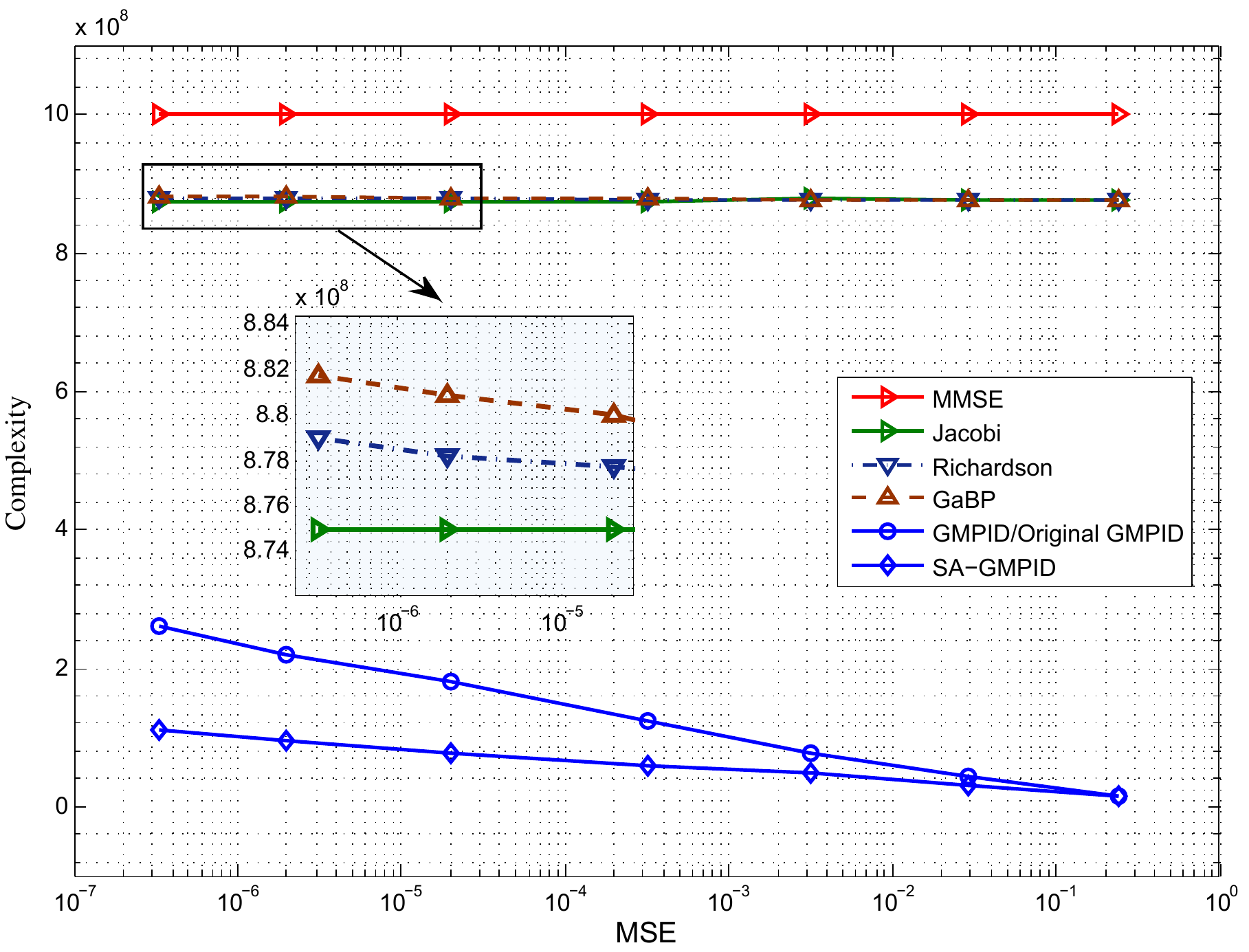}\\\vspace{-0.2cm}
  \caption{Complexity comparison between the MMSE detector, Jacobi algorithm, GaBP algorithm, Richardson algorithm, GMPID and SA-GMPID for the MU-MIMO system that $K=500$, $M=3500$ and $\beta=1/7$. }\label{f9}\vspace{-0.2cm}
\end{figure}

\begin{figure*}[ht]
  \centering
  \includegraphics[width=16.8cm]{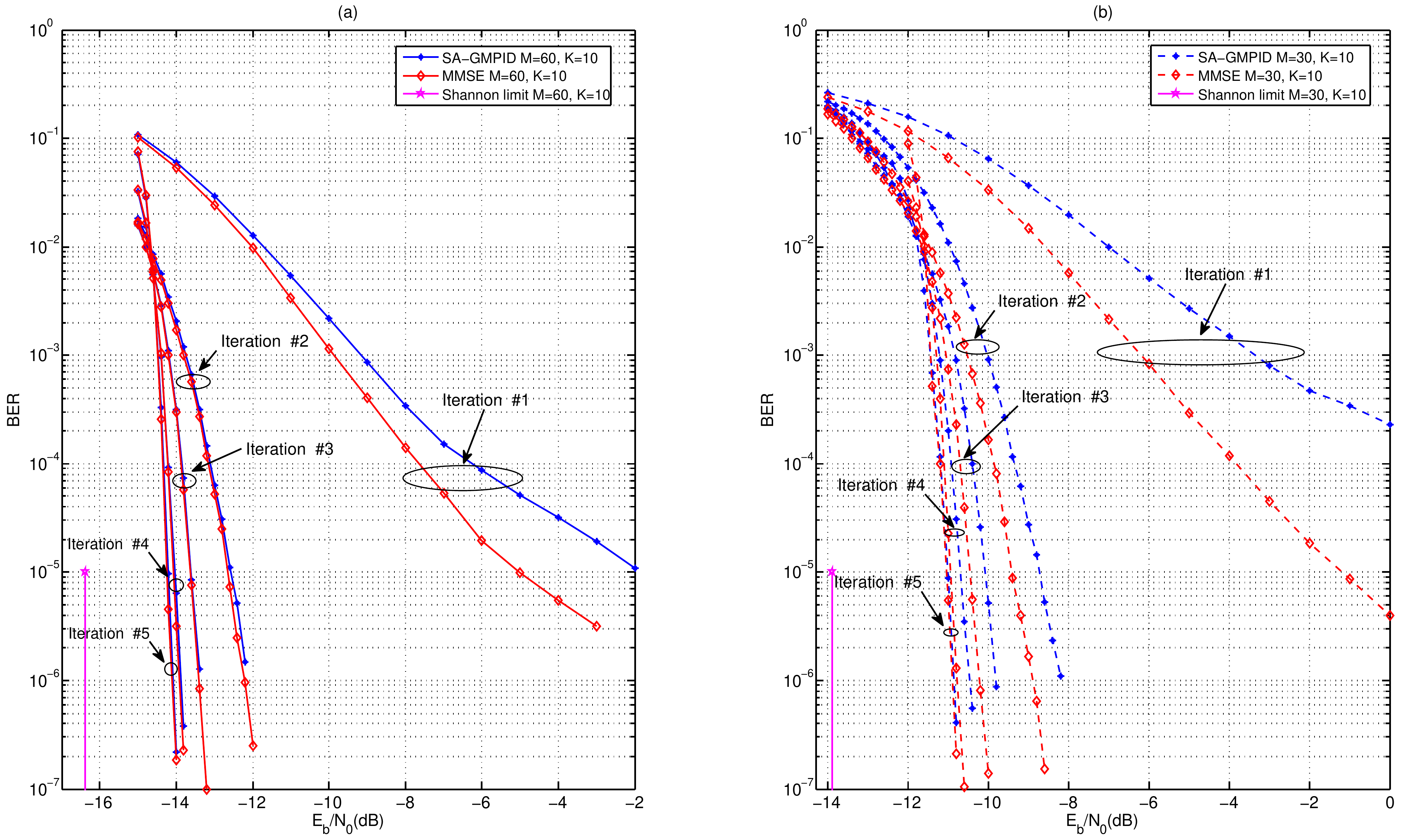}\\\vspace{-0.0cm}
  \caption{BER performances of SA-GMPID and MMSE detection with Gaussian approximation for the discrete MU-MIMO communication systems, where in sub-figure (a), $K=10$, $M=60$ and $\beta=1/6$, and in sub-figure (b), $K=10$, $M=30$ and $\beta=1/3$. Each user is encoded by a Turbo Hadamard code (3 component codes, Hadamard order: 5, spread length: 4) with code rate 0.01452 bits/symbol and code length $2.82\times 10^5$, and then 10bits superposition coded modulation is employed for the Turbo Hadamard code. Transmitting length of each user is $2.82\times 10^4$. The rate of each user is 0.1452 bits/symbol, and the system sum rate is 1.452 bits per channel use. The Shannon $E_b/N_0$ limit of system (a) is -16.37dB, and system (b) is -13.88dB. }\label{f10}\vspace{-0.0cm}
\end{figure*}

\section{Conclusion}
GMPID is a low-complexity multi-user detection algorithm in which the means and variances are transmitted between the variable nodes and sum nodes. The convergence of GMPID is analysed in this paper. It is proved that the variances of GMPID converge to the MSE of MMSE detection. Two sufficient conditions that the GMPID converges to the MMSE detection are presented. As GMPID does not converge when $\beta\!\geq\!(\sqrt{2}\!-\!1)^2$, the SA-GMPID algorithm is proposed, which is proved to converge to the MMSE detection for any $\beta\!<\!1$ (underloaded system) with a faster convergence speed and has no higher complexity as the GMPID. Numerical results are provided to verify the theoretical results. Interestingly, our simulations show that the proposed theorems even work for MU-MIMO systems with a small number of antennas and users (e.g., tens or less).

When $\beta>1$ (overloaded system), it can be shown that although the GMPID may converge to the MMSE detection very quickly, the system performance may be very poor. However, overloaded MU-MIMO will be the future trend of wireless communication networks. Therefore, we will focus on the case of $\beta>1$ in our future work.

\end{document}